\documentclass[%
 reprint,
showpacs,preprintnumbers,
 amsmath,amssymb,
aps,
pra,
floatfix,
]{revtex4-1}
\usepackage{amsmath,amsfonts,amsthm} 
\usepackage{dsfont}
\usepackage{braket}
\usepackage{graphicx} 
\usepackage{epstopdf}
\usepackage{dcolumn} 
\usepackage{bm} 
\usepackage{float}
\usepackage[caption=false]{subfig}
\usepackage{color}
\usepackage{float}
\begin{document}

\preprint{APS/123-QED}

\title{Detection of out-of-time-order correlators and information scrambling in cold atoms: Ladder-XX model}
\author{Ceren B. Da\u{g}}
\email{cbdag@umich.edu}
\author{L.-M. Duan}
\affiliation{Department of Physics, University of Michigan, Ann Arbor, Michigan 48109, USA}
\date{\today}

\begin{abstract}
Out-of-time-order correlators (OTOC), recently being the center of discussion on quantum chaos, are a tool to understand the information scrambling in different phases of quantum many-body systems. We propose a disordered ladder spin model, XX-ladder, which can be designed in a scalable cold atom setup to detect OTOC with a novel sign reversal protocol for the evolution backward in time. We study both the clean and disordered limits of XX-ladder and characterize different phases (ergodic-MBL) of the model based on the decay properties of OTOCs. Emergent effective lightcone shows sublinear behaviour, while the butterfly cones drastically differ from the lightcone via demonstrating superlinear behaviour. Based on our results, one can observe how the information scrambling changes in the transition from well-studied 1D spin models to unexplored 2D spin models in a local setting.
\end{abstract}


\maketitle


Information scrambling has drawn much attention in the last years, not only in gravitational theories to study the information properties of black holes \cite{1126-6708-2008-10-065,kitaev, 2014JHEP...03..067S,Maldacena2016}, but also in quantum many-body physics \cite{doi:10.1002/andp.201600332,2017PhRvB..95f0201S,2017PhRvB..95e4201H,doi:10.1002/andp.201600318,2016arXiv160801914F,PhysRevLett.121.016801,PhysRevB.96.020406,2018arXiv180506895L,2017NJPh...19f3001B,2018arXiv180200801X,2018arXiv180200801X}. Even though the initial interest in scrambling was to study quantum chaos in models with gravity duals, information scrambling is, first, not limited to systems with duals, second, provides an understanding about the dynamics of any generic many-body system. Besides being a complementary approach to level-statistics \cite{doi:10.1063/1.1703773} in the context of quantum chaos, the way that the systems scramble information in time can dynamically reveal the properties of a Hamiltonian in an experiment. The tool to measure the information scrambling is a correlation function, the so-called out-of-time-order correlator (OTOC). The physics that OTOC captures is the growth of the commutator of two operators in time and this growth can be characterized by,
\begin{eqnarray}
C_{i}^{\beta} (t) &=& -\frac{1}{Z} \text{Tr} \left\lbrace e^{-\beta H} \left[A_i(t),B_{j=0}\right]^2 \right\rbrace,
\label{OTOCGeneral} 
\end{eqnarray}
for a system with a finite inverse temperature $\beta$. Here $i$ denotes a site in the lattice, $j=0$ is the first lattice site, $A_i(t)$ and $B_{j=0}$ are local hermitian operators for their corresponding sites and $Z$ is the partition function. The local observables of two sites at a distance initially commute, but the interactions lead the system to become more correlated in time, and the build-up of the correlations between sites-at-a-distance starts to be seen in the Heisenberg operators that no longer commute. Therefore, the initially localized operators spread across the space dimension and become as nonlocal as possible around the scrambling time. OTOCs are sensitive to conserved quantities \cite{doi:10.1002/andp.201600332,PhysRevX.7.031011,PhysRevB.97.144304}, revealing the (non)integrability of the system; they also show the signatures of localized phases \cite{doi:10.1002/andp.201600332,2017PhRvB..95f0201S,2017PhRvB..95e4201H,doi:10.1002/andp.201600318,2016arXiv160801914F}, equilibrium \cite{2019arXiv190205041D} and dynamical phase transitions \cite{PhysRevLett.121.016801}, chaotic properties of thermal systems \cite{Maldacena2016,2017JHEP...10..138H,2017NJPh...19f3001B}, e.g. exponential decay in OTOC and finally the (non-)locality and information transport of the Hamiltonian via emergent lightcones \cite{PhysRevB.96.020406,2018arXiv180506895L,2017NJPh...19f3001B,2018arXiv180200801X}. All these theoretical discoveries on OTOCs call for experimental proposals and experiments in order to probe and eventually utilize scrambling. 

To date, there have been a number of experimental proposals \cite{PhysRevA.94.040302,2016arXiv160701801Y,PhysRevA.94.062329,2017NJPh...19f3001B,PhysRevLett.120.070501} and realizations \cite{PhysRevX.7.031011, garttner2017measuring, 2018arXiv180602807L} on scrambling detection. In this paper, one of our aims is to come up with the simplest possible cold atom setup that shows a wide range of diverse scrambling phenomena and could pave the way to the scalable OTOC measurements of non-integrable spin systems. The cold-atom setup is a realistic candidate to probe OTOC, mainly due to scalability and its weak coupling to the environment \cite{RevModPhys.80.885,RevModPhys.83.863}. Information scrambling could be induced by environment effects as well, and therefore it is important to differentiate the scrambling due to correlation built-up via many-body interactions in an experiment \cite{2018arXiv180602807L}. The scalability of cold atoms could be utilized to increase the size and hence the duration of transient effects in OTOC by delaying the saturation stage. The most crucial step of OTOC measurement is the evolution backward in time. We propose a novel sign reversal mechanism as an alternative to existing approaches. The convential solution to reverse the sign of a cold-atom Hamiltonian is to utilize Feshbach resonances \cite{Pethick2002,2017NJPh...19f3001B}. We will show that a sequence of single-spin gates can be performed via fast laser pulses \cite{PhysRevLett.105.090502, 2014NatSR...4E5867L} to measure the OTOCs.

In the first section, we explain our model and its cold-atom setup. Then we systematically study the level-statistics and scrambling properties of XX-ladder both with and without disorder. In the final part, we layout the scrambling detection with the preparation of realistic random states.

\section{The Ladder-XX Model} 

Ladder spin models have been studied to explore their critical phenomena \cite{PhysRevB.47.3196,PhysRevLett.69.2419,2007AdPhy..56..465B} and entanglement properties \cite{PhysRevA.79.012331}. They are seen as useful intermediate models to understand the magnetic properties of materials while increasing the dimension from $d$ to $d+1$ \cite{Dagotto618}. There are also natural cuprate compounds that are modelled by ladder spin models at $d=1$ \cite{2007AdPhy..56..465B} and they have been considered as candidate models to explain high-$\text{T}_{\text{c}}$ superconductivity \cite{0034-4885-62-11-202}. More recently ladder-spin models are studied in the context of transport \cite{PhysRevLett.110.070602}. We set our chaotic ladder model as the \emph{ladder-XX} model because of its simplicity in cold atom realization, 
\begin{eqnarray} \label{EqHamiltonian} 
H &=& \sum_{j=1,2}\sum_{i=1}^{L-1} J_{\parallel} \left(\sigma_{j,i}^x \sigma_{j,i+1}^x + \sigma_{j,i}^y \sigma_{j,i+1}^y \right)\\ 
&+& \sum_{i=1}^{L} J_{\perp} \left(\sigma_{1,i}^x \sigma_{2,i}^x + \sigma_{1,i}^y \sigma_{2,i}^y \right) + \sum_{i=1}^{L} h_i \left(\sigma^z_{1,i}+ \sigma^z_{2,i}\right), \notag 
\end{eqnarray}
with random disorder $h_i$ which is drawn from a uniform distribution with disorder strength of $[-h,h]$. $\sigma^{x,y,z}$ are Pauli matrices for the spin$-1/2$ system, $J_{\parallel}$ is the intra-chain hopping coefficient and $J_{\perp}$ is the rung hopping coefficient. $L$ is the system size for a single-chain and we go up to $L=8$ in our numerical analysis with exact diagonalization. 

The ladder-XX model could be realized at the hard-core boson limit of the Bose-Hubbard model \cite{PhysRevLett.81.3108,2002Natur.415...39G}. At the hard-core boson limit, with $U \rightarrow \infty$ and non-integer filling factor that implies every site has either 0 or 1 boson, 
we end up with a superfluid Hamiltonian $ H_{U\rightarrow \infty} = -t_{\parallel} \sum_{i,i+1} \left(a_i^{\dagger} a_{i+1} + \text{h.c.}\right) - \sum_i \mu_i a_i^{\dagger} a_i$, that can easily be mapped to XX-chain via mapping the annihilation operator to the spin lowering operator $a \rightarrow \sigma^-$ and creation operator to raising operator $a^{\dagger} \rightarrow \sigma^+$. The mapping leads us to have $J_{\parallel} = 2 t_{\parallel}$, $J_{\perp} = 2 t_{\perp}$ and the random chemical potential is mapped to random magnetic field strengths $\mu_i = h_i$ via $a_i^{\dagger} a_i - 1/2 \rightarrow \sigma^z$. Therefore, we can recover Hamiltonian Eq. \ref{EqHamiltonian} with two interacting Bose-Hubbard chains exposed to random chemical potential in the hard-core boson limit. The boson state vectors correspond to either spin down $\Ket{\downarrow}$ or spin up $\Ket{\uparrow}$ in the ladder-XX model. Since the filling factor is fixed in the cold atom scheme, the corresponding case in our spin model (Eq. \ref{EqHamiltonian}) has fixed total spin $S_z$. We set the filling factor $f = 0.5$ and the OTOC of the system is studied at the subsector $S_z = 0$. 

We utilize superlattices to create random disorder in the Bose-Hubbard chains \cite{PhysRevLett.91.080403,PhysRevLett.98.130404} and to let two chains interact with each other. For the latter, we create a double well potential via choosing the laser frequencies as $k$ and $2k$ in the y-direction with a phase difference between them $\phi$, e.g. $V_y(y) = V_{1y} \sin^2 \left( k_y y \right) + V_{2y} \sin^2 \left(2 k_y y + \phi \right)$, assuming $V_{1y} \sim V_{2y}$ so that the bosons can be trapped in double well potential. For the random disorder, we interfere two optical fields with incommensurate frequencies, e.g. $V_x (x) = V_{1x} \sin^2 \left(k_{1x} x \right) + V_{2x} \sin^2 \left(k_{2x} x \right)$, where $k_{1x}/k_{2x} \in \mathbb{R} / \mathbb{Q}$ for both of the chains. When $V_{2x} \ll V_{1x}$, the disorder lattice can simulate the true random potential \cite{PhysRevLett.91.080403,PhysRevLett.98.130404}. One can tune the hopping coefficients $J_{\parallel}$ and $J_{\perp}$ in the ladder-XX model through the laser amplitudes and frequencies \cite{PhysRevLett.81.3108}; and thus access different OTOC behaviours with the simulation time of $t \propto 1/J_{\parallel} \propto$ 1-10 ms in laboratory. Therefore, the measurement time of OTOC is in the limits of cold atom experiments \cite{2002quant.ph..7011J}.


\section{The OTOC properties and level statistics} 

For a spin system Eq. \ref{OTOCGeneral} can be recast to the OTOC, by first setting the temperature infinite, $\beta \rightarrow 0$ and then noting that,
\begin{eqnarray}
F_i^{\text{ex}}(t) = 1-\frac{C_i^0(t)}{2 N},\label{OTOC}
\end{eqnarray}
where $ C^0_{i}(t)= \| \left[\sigma^z_i(t),\sigma^z_1 \right]  \|_{F}^2$. Since the Pauli matrices are hermitian, norm-2 (Frobenius norm) could be utilized to rewrite Eq. \ref{OTOCGeneral}. $N$ is the dimension of the Hilbert space and the superscript $\text{ex}$ stands for the exact value of the out-of-time-order correlator. Eq. \ref{OTOC} is measurable given a $\beta=0$ initial state is prepared. In general, calculating an expectation value with respect to the infinite temperature state requires averaging over all eigenstates. However, we can approximate the OTOC Eq. \ref{OTOC} with smaller number of states,
\begin{eqnarray}
F_{i}^{\sim}(t) &=& \sum_{j} \Bra{\psi_j} \sigma^z_{i}(t) \sigma^z_{1} \sigma^z_{i}(t) \sigma^z_{1} \Ket{\psi_j},\label{approxOTOC}
\end{eqnarray}
where $\Ket{\psi_j}$ denotes a pure random initial state (or a mixture of random initial states) drawn from the Haar measure \cite{PhysRevB.96.020406}. Haar random states are typically maximally-entangled states within a small error \cite{Hayden2006}. The error of approximating a $\beta=0$ initial state is exponentially suppressed as the Hilbert space increases via typicality arguments \cite{2017AnP...52900350L,popescu2006entanglement}. This procedure is numerically less expensive compared to other methods for preparing the initial state at $\beta=0$, even though the Haar random states are hard to generate experimentally \cite{Emerson2098}. The results presented in this paper are based on averaging over more than one random initial state to obtain OTOC as precise as possible (App.~B for error bounds). 

\begin{figure}
\centerline{\includegraphics[width=0.45\textwidth]{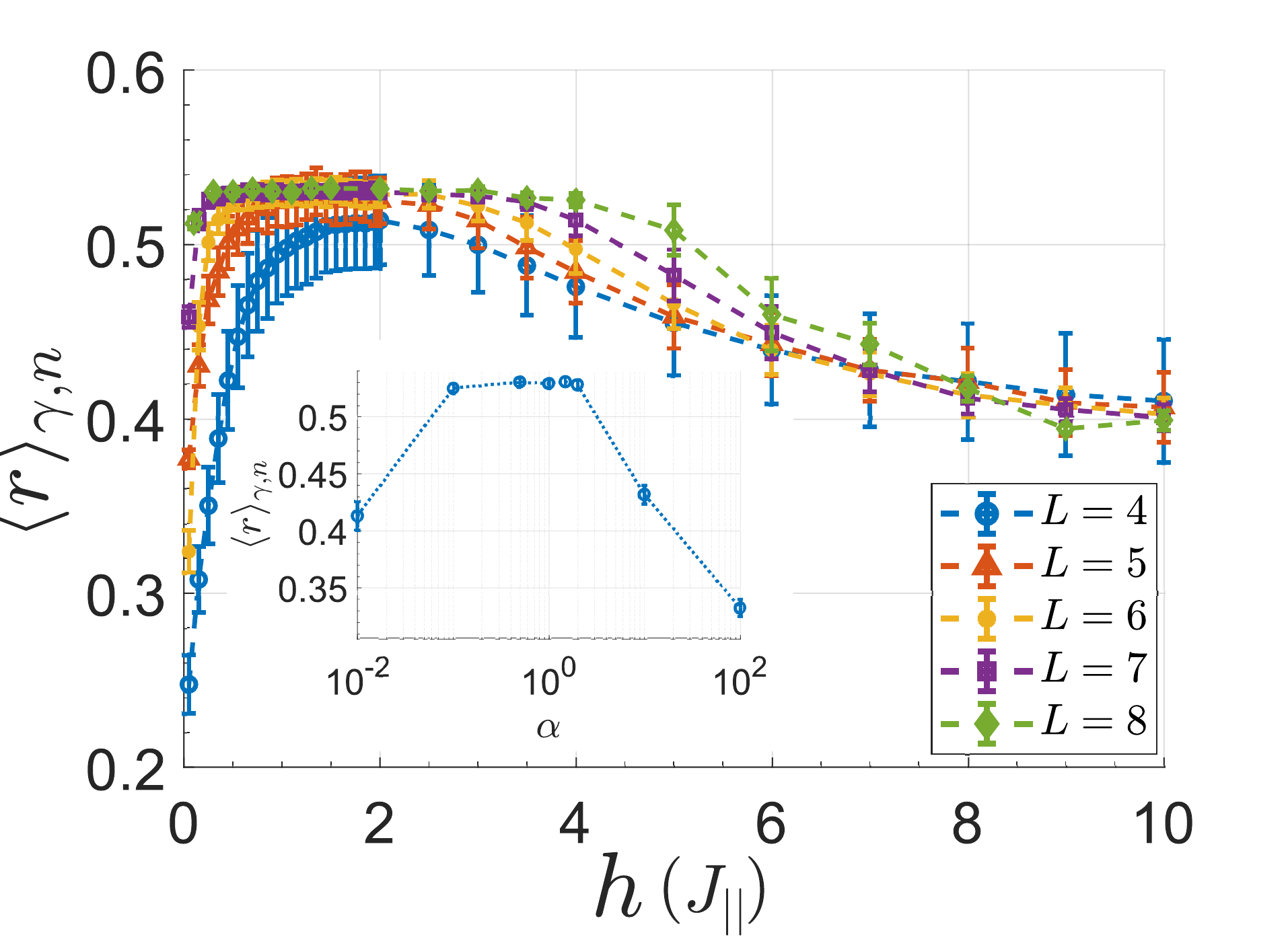}}
\caption{The average ratio of level spacings $\Braket{r_{\gamma}^n}_{\gamma,n}$ with respect to disorder strength  $h$. Coupling strengths are set to $J_{\perp} = J_{\parallel}$ and $\Braket{r_{\gamma}^n}_{\gamma,n}$ is averaged over $5\times 10^3$ to $10$ random realizations for single-chain sizes ranging between $L=4-8$. Inset: $\Braket{r_{\gamma}^n}_{\gamma,n}$ at $h = 1~[J_{||}]$ with respect to rung interaction strength $\alpha$ where $J_{\perp} = \alpha J_{\parallel}$ for  $L=7$. 
} \label{SuppFig13}
\end{figure}

When a generalized form of Jordan-Wigner transformation \cite{PhysRevB.50.6233} is applied, XX-ladder can be shown to be interacting in the spinless fermion representation. Therefore we expect to see ergodic to many-body localized (MBL) phase transition in this model \cite{PhysRevB.75.155111,PhysRevB.82.174411}. A common way to determine if a quantum system is chaotic is via the energy level statistics \cite{doi:10.1063/1.1703773,PhysRevB.75.155111,PhysRevB.82.174411,2016AdPhy..65..239D}. Energy level spacings are $\delta_{\gamma}^{n} = |E_{\gamma}^{n}-E_{\gamma}^{n-1}|$ where $E_{\gamma}^{n}$ is the corresponding energy of the many-body eigenstate $n$ in a Hamiltonian of disorder realization $\gamma$. Each $\gamma$ represents a different set of random disorder $h_i$ drawn from uniform distribution. Then we can calculate the ratio of adjacent gaps as $r_{\gamma}^n = \text{min}\left( \delta_{\gamma}^n,\delta_{\gamma}^{n+1} \right)/\text{max}\left( \delta_{\gamma}^n,\delta_{\gamma}^{n+1} \right)$ as the indicator of the level-statistics \cite{PhysRevB.75.155111,PhysRevB.82.174411}: $r_{\gamma}^n \sim 0.53$ and $r_{\gamma}^n \sim 0.39$ are representative of Wigner-Dyson and Poisson statistics, respectively. If the distribution of the energy level spacings follows Wigner-Dyson statistics through a GOE (generalized orthogonal ensemble) distribution, the model shows ergodic behaviour, whereas Poisson statistics imply a localized phase \cite{doi:10.1063/1.1703773,2016AdPhy..65..239D}. Fig. \ref{SuppFig13} shows the average ratio values $\Braket{r_{\gamma}^n}_{\gamma,n}$ varying between random field strengths of $h = 0-10$ for different system sizes ranging between $L=4$ and $L=8$ when they are averaged over $5\times 10^3$ to $10$ different random samples. The average of $r_{\gamma}^n$ over a set of different Hamiltonians $H_{\gamma}$ and eigenstates $n$, converges to $\Braket{r_{\gamma}^n}_{\gamma,n} \sim 0.53$ in the presence of small disorder strength $h \lesssim 3~[J_{||}]$, hence implying an ergodic phase. As $h \gtrsim 9~[J_{||}]$, we observe $\Braket{r_{\gamma}^n}_{\gamma,n} \sim 0.39$ that indicates a many-body localized (MBL) phase.

\begin{figure}
\centerline{\includegraphics[width=0.45\textwidth]{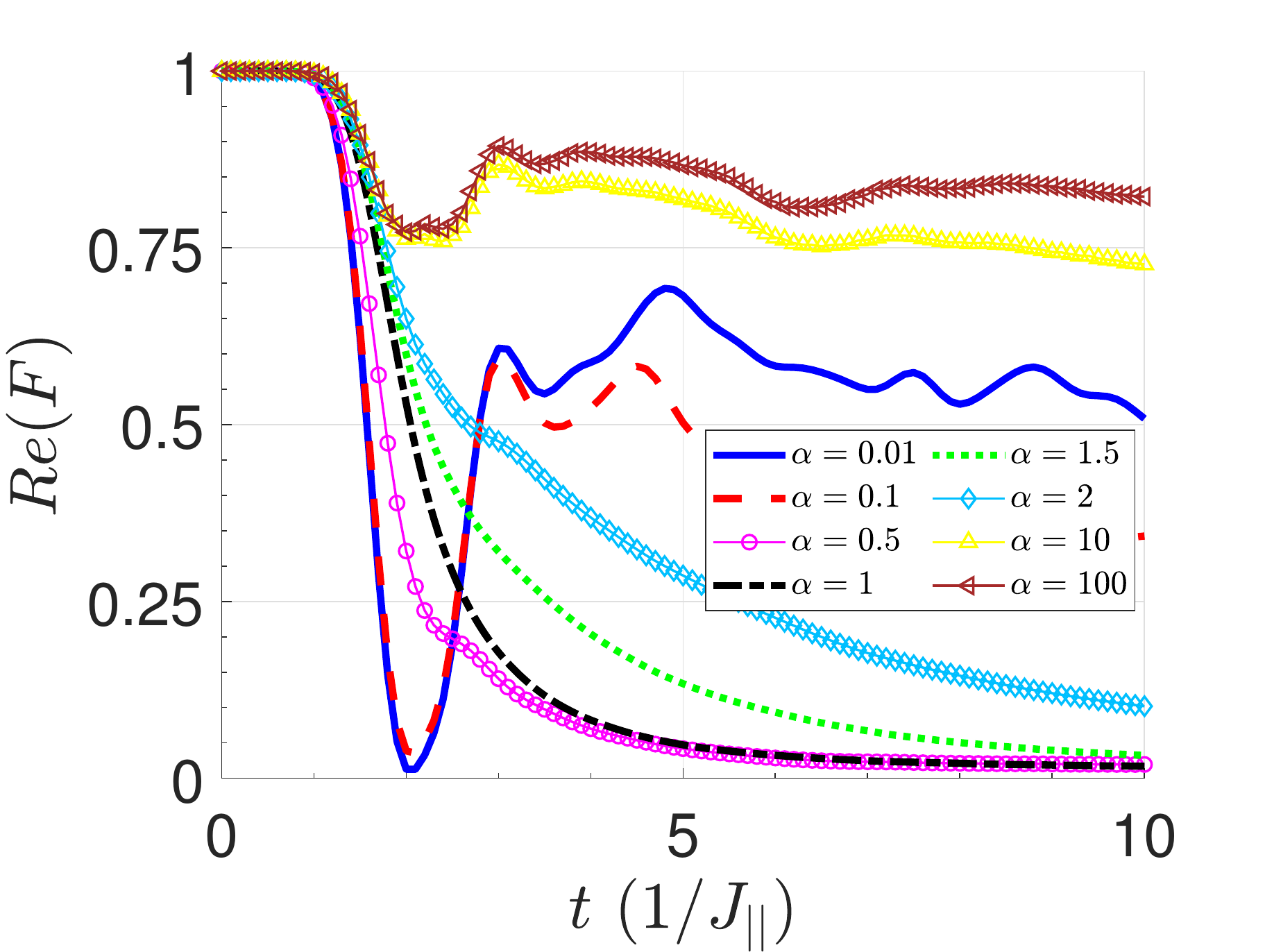}}
\caption{The OTOC of the ladder-XX model at $h = 1~[J_{||}]$ between two distant operators $\sigma^z_1$ and $\sigma^z_7$ in the first chain with respect to $\alpha$ for $L=7$. $\alpha \sim 1$ corresponds to the interacting limit, whereas the cases $\alpha \ll 1$ and $\alpha \gg 1$ are integrable limits of the ladder-XX model. The OTOC is averaged over 100 different random samples. The plot shows the mean values, see App.~A for the error bars on the curves.} \label{Fig2}
\end{figure}

Fig. \ref{Fig2} shows how OTOC between $\sigma^z_1$ and $\sigma^z_7$ for $L=7$ chain changes with respect to the rung interaction strength. At the limit of $\alpha = J_{\perp}/J_{\parallel} \rightarrow 0$, the system converges to two independent XX-chain with random disorder. Whereas the opposite limit of $\alpha \rightarrow \infty$ implies a dimer phase as another integrable limit of XX-ladder. In both cases, the corresponding fermion representation becomes non-interacting, hence points to single-particle dynamics with Anderson localization \cite{2016arXiv160801914F,doi:10.1002/andp.201600332}. We see a permanent revival after a decay and larger oscillations in OTOC as observed in integrable systems \cite{2017JHEP...10..138H,PhysRevX.7.031011}. In addition, the average level spacing ratio $\Braket{r_{\gamma}^n}_{\gamma,n}$ decreases from $\sim 0.53$ to $\sim 0.39$, thus demonstrating level statistics for integrable systems (see inset of Fig. \ref{SuppFig13}). We note that the OTOC for $\alpha \rightarrow \infty$ scrambles less than the OTOC for $\alpha \rightarrow 0$ with a small initial decay, since the model also becomes weakly-coupled throughout the x-dimension in this limit. The OTOC decays rapidly in the interacting limit around $\alpha \sim 1$ and saturates at $F(t\rightarrow \infty) \sim 0$ while showing GOE distribution with $\Braket{r_{\gamma}^n}_{\gamma,n} \sim 0.53$ and hence quantum chaos in its energy levels. We set $\alpha = 1$ for the rest of our paper and study the interacting limit.

\begin{figure}
\centering
\subfloat[]{\label{Fig3a}\includegraphics[width=0.24\textwidth]{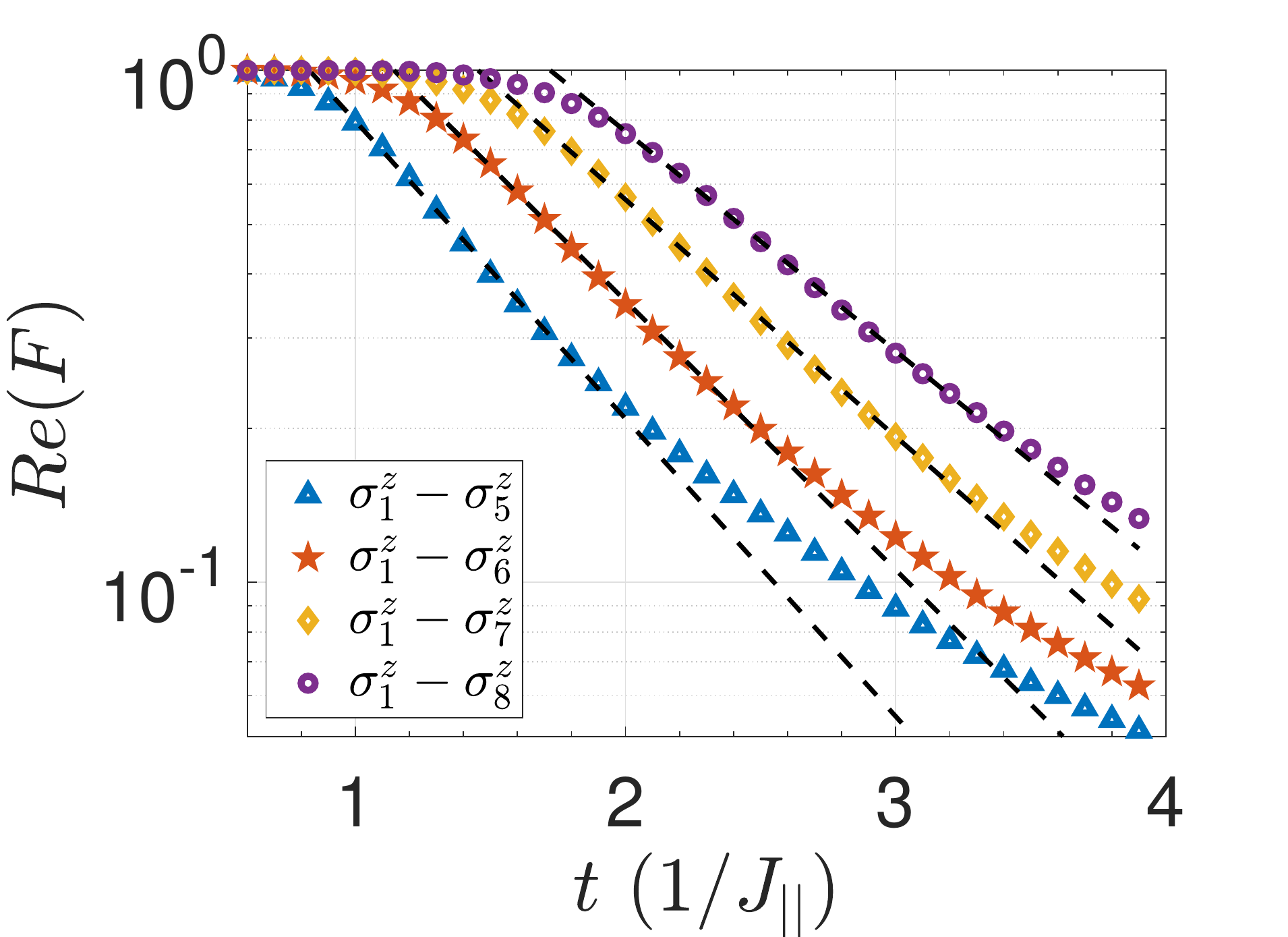}}\hfill \subfloat[]{\label{Fig3b}\includegraphics[width=0.24\textwidth]{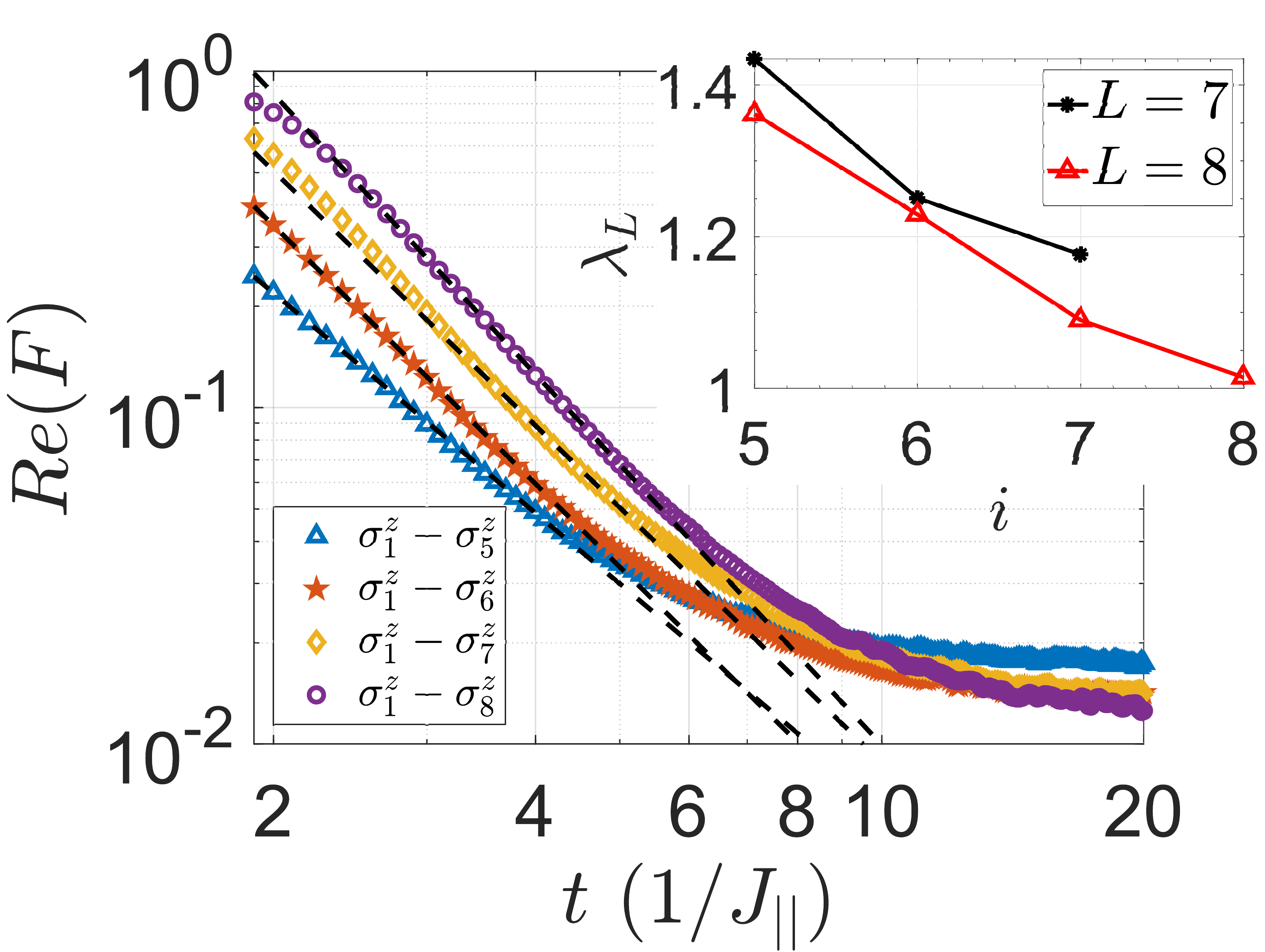}}\hfill \subfloat[]{\label{Fig3c}\includegraphics[width=0.24\textwidth]{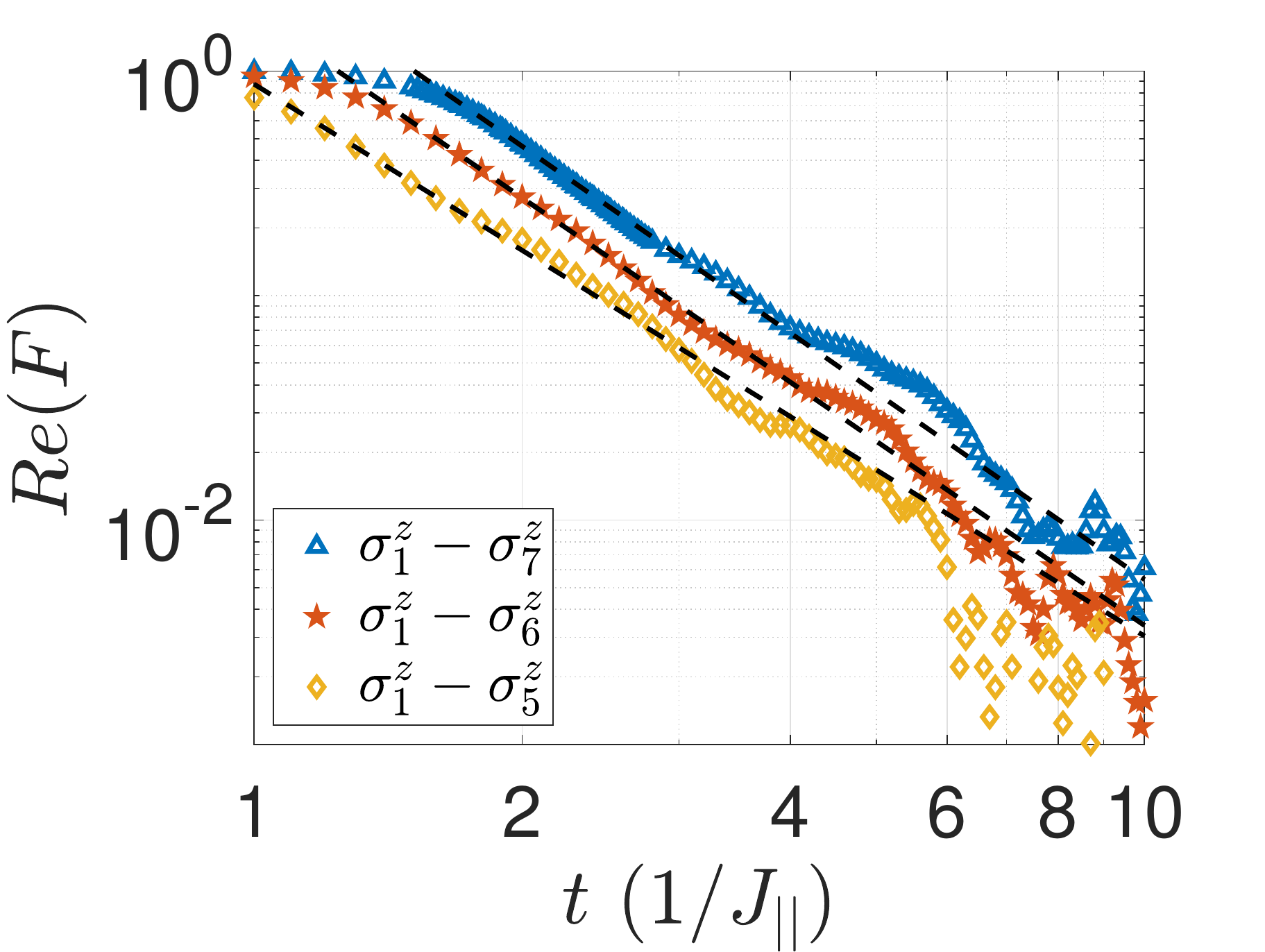}}\hfill \subfloat[]{\label{Fig3d}\includegraphics[width=0.24\textwidth]{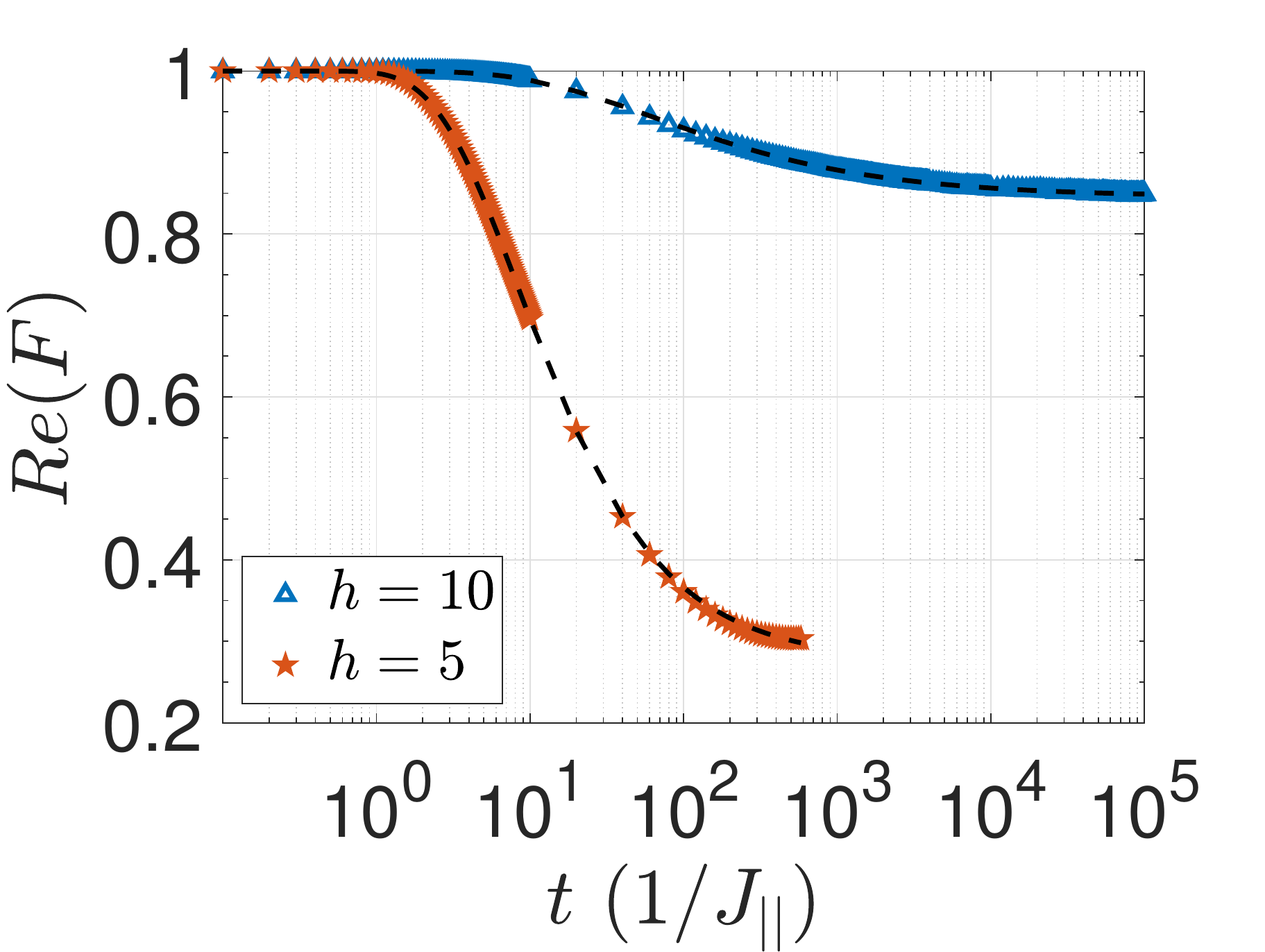}}
\caption{(a) The exponential and (b) power-law decay of OTOCs for $\sigma_1^z$ with $\sigma_5^z$ (blue-triangles), $\sigma_6^z$ (red-pentagrams), $\sigma_7^z$ (orange-diamonds) and $\sigma_8^z$ (purple-circles) observables in a system size of $L=8$. The inset in (b) shows the Lyapunov-like exponent extracted from exponential fitting for both $L=7$ (black-asterisks) and $L=8$ (red-triangles). (c) No disorder case: Only power-law decay of OTOC for $\sigma_1^z$ with $\sigma_5^z$ (orange-diamonds), $\sigma_6^z$ (red-pentagrams) and $\sigma_7^z$ (blue-triangles) observables when $L=7$ and $h=0$. (d) Crossover region with $h=5~[J_{||}]$ (red-pentagrams) and MBL with $h=10$ (blue-triangles) for observables $\sigma_1^z-\sigma_7^z$ with $L=7$.}
\label{Fig6}
\end{figure}

The chaotic regime of the ladder-XX model ($h=1~[J_{||}]$) demonstrates a brief interval of exponential decay in early-time dynamics (Fig. \ref{Fig3a}), followed by power-law tails (Fig. \ref{Fig3b}) before entering into the saturation regime. The inset in Fig. \ref{Fig3b} shows the Lyapunov-like exponents extracted from the data both for $L=8$ and $L=7$ (App.~C) when we fit $\text{Re}(F) = a \exp (-\lambda t)$ to the data, where $a$ is a constant. Quantum chaotic models are expected to scramble the information fast and hence show exponential decay of OTOC \cite{Maldacena2016} before the saturation. Exponential decay is a transient feature of systems with finite size and bounded operators \cite{2017PhRvB..96f0301K}, a result we observe in Fig. \ref{Fig3a}. The Bose-Hubbard model \cite{2017NJPh...19f3001B} and time-dependent systems \cite{doi:10.1002/andp.201600332,2018arXiv180401545R} were shown to decay exponentially, whereas it is numerically hard to show the exponential decay in time-independent quantum chaotic spin chains, e.g. disordered Heisenberg model \cite{doi:10.1002/andp.201600332}. In fact, the transient exponential decay turns into power-law decay $\text{Re}(F) = a t^{-b}$ in Fig. \ref{Fig3b} for the ladder-XX model, thus reminding us of the quasi-exponential generic form put forward by \cite{2018arXiv180200801X}. When there is no disorder, Fig. \ref{Fig3c}, a decay with power-law trend is observed. There are significantly larger oscillations around the saturation value in the clean limit, however in both clean and disordered cases, the scrambling time is approximately the same. For a comparison, the power-law exponents for disordered and clean cases are $b=2.65$ and $b=2.76$, respectively for the observables $\sigma_1^z-\sigma_7^z$ in a system with $L=7$. The ladder-XX model has energy and spin conservation, similar to Heisenberg model where OTOC has been observed to be sensitive to conserved quantities and show power-law decay \cite{doi:10.1002/andp.201600332}. In addition to that, XX-ladder has invariant subspaces that show ballistic transport but are not associated with local conserved quantities at the same time, hence the energy levels still show quantum chaos \cite{PhysRevLett.110.070602}. When random disorder is introduced, these invariant subspaces can support Anderson localized eigenstates regardless of the disorder strength \cite{2018arXiv181107903I}. We first conclude that the invariant subspaces do not change the power-law decay, however affect the saturation value of OTOC. Figs. \ref{Fig3b}-\ref{Fig3c} show that the saturation value is much higher both in disordered $F(t\rightarrow \infty) > 10^{-2}$ and clean $F(t\rightarrow \infty) > 10^{-3}$ limits, compared to other models such as Heisenberg and transverse-field Ising models of similar sizes $F(t\rightarrow \infty) \sim 10^{-5}$ \cite{doi:10.1002/andp.201600332}. Further, we notice that the saturation value of OTOC becomes even larger when the disorder is introduced. 
Even though the disorder clearly resolves the degeneracies caused by symmetries, the disordered system scrambles less than the clean system. Thus, we point to Griffiths rare-region effects \cite{doi:10.1002/andp.201600326} that might also be responsible for turning exponential decay in early time into power-law later in time.

The decay becomes even slower as we increase the disorder strength $h$, Fig. \ref{Fig3d}. The system shows no scrambling for a time interval of $t \sim 10 [1/J_{||}]$ when $h=10~[J_{||}]$ and differs from OTOC at $h=5~[J_{||}]$ that is at the crossover region in Fig. \ref{SuppFig13}. Even though for short times it looks like Anderson localization, simulation over long times reveals an MBL-like decay by showing a clear signature of logarithmic decay at intermediate times for both $h=5$ and $h=10$. By slightly modifying the general form given in Ref. \cite{2017PhRvB..95e4201H} for logarithmic MBL decays, we find that the decay profiles in Fig. \ref{Fig3d} could be fitted to $Re(F) = 1-a\exp \left(-bt^c\right)$, where the parameter $a$ determines the saturation value, and $c<0$ for OTOC to decay as $t \rightarrow \infty$ and $F=1$ as $t \rightarrow 0$. Similarly this form reduces to logarithmic decay, $Re(F) \sim 1-\frac{a}{e}+\frac{a\times c}{e} \log \left(b^{1/c}t\right)$ for $b^{1/c}t \sim 1$. The fit parameters read $a=0.725$, $b=5.727$, $c=-0.812$ for $h=5$ and $a=0.154$, $b=8.661$, $c=-0.519$ for $h=10~[J_{||}]$. Therefore, the logarithmic decay is valid around $t\sim 10~[1/J_{||}]$ and $t=10^2~[1/J_{||}]$ for $h=5~[J_{||}]$ and $h=10~[J_{||}]$, respectively. One can further see that Anderson localization lies in the limit $|c| \rightarrow 0$, which implies logarithmic decay should happen when $t \rightarrow \infty$, meaning that the OTOC does not decay at all. As a result, we demonstrate that there could be intermediate cases where the OTOC does not decay to zero, but to finite nonzero values in the MBL phase, which is possibly related to atypical eigenstates in XX-ladders \cite{2018arXiv181107903I}.

\begin{figure}
\centerline{\includegraphics[width=0.45\textwidth]{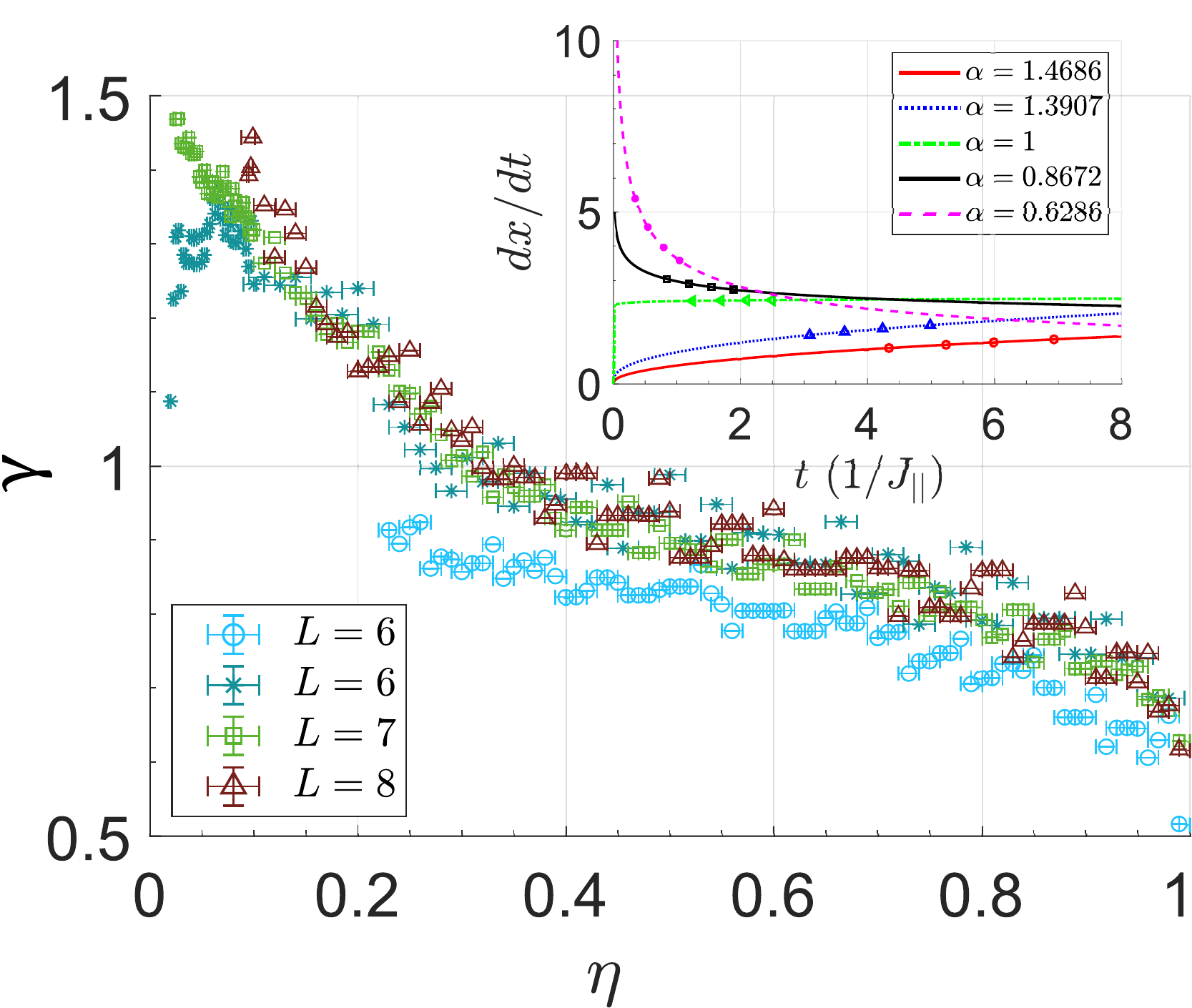}}
\caption{The dynamical exponent $\gamma$ with respect to the OTOC contour values $\eta$ extracted from analyzing data sets for $L=6$ with observables from $\sigma_2^z$ to $\sigma_6^z$ (light blue-circles), with observables from $\sigma_4^z$ to $\sigma_6^z$ (dark blue-stars), $L=7$ from $\sigma_4^z$ to $\sigma_7^z$ (green-squares) and $L=8$ from $\sigma_4^z$ to $\sigma_8^z$ (red-triangles) for a random disorder strength of $h=1$. We averaged the data over $2\times 10^2$, $1\times 10^2$, $1\times 10^2$ and $1\times 10^1$ times for first two $L=6$, $L=7$ and $L=8$ system sizes, respectively. Inset: The rates of the sublinear, linear and superlinear wavefronts for a system size of $L=7$. The markers are the data points, while the lines are the differentiation of the wavefront curves.
} \label{Fig7}
\end{figure}
\begin{figure}
\centering
\subfloat[]{\label{Fig6a}\includegraphics[width=0.24\textwidth]{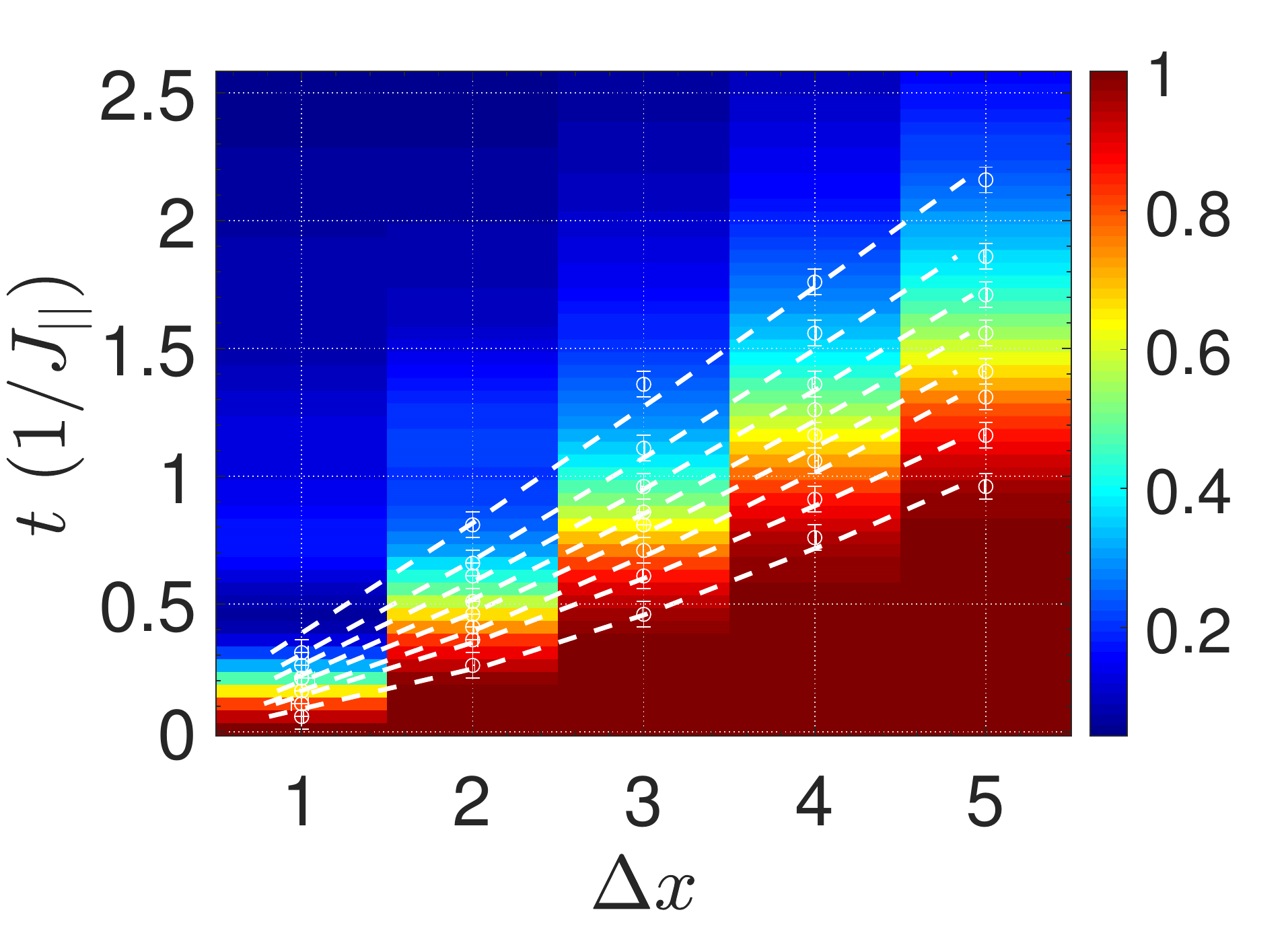}}\hfill \subfloat[]{\label{Fig6b}\includegraphics[width=0.24\textwidth]{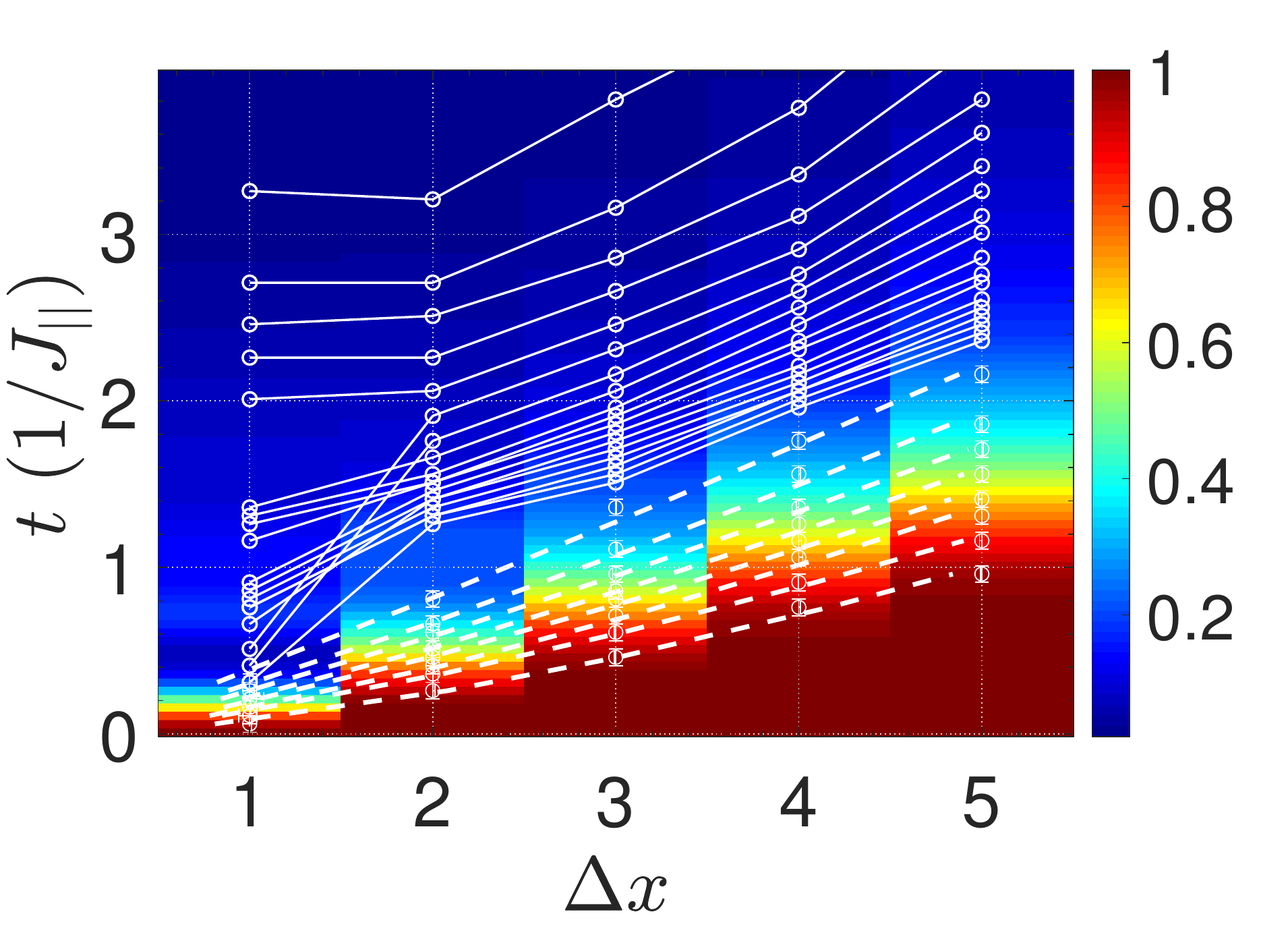}}
\caption{A demonstration of wavefronts for a system size of (a) $L=8$ and (b) $L=6$, where x-axis and y-axis are the distance and time, respectively. (a) The fitted wavefronts change from sublinear to superlinear in time between the displacements $\Delta x=3$ and $\Delta x=7$ units. (b) Only the sublinear wavefronts are fitted between $\Delta x=1$ and $\Delta x=5$ units (dotted lines), while the solid lines show irregular wavefronts appearing later in time.}
\label{Fig5}
\end{figure}

In a lightcone figure (Fig. \ref{Fig5}), each point has a set of discrete space $x$ and time $t$ coordinates, where the space dimension is emergent due to the nearest-neighbor couplings and defined as the distances between lattice sites in the lower leg of the ladder. The value of a point is OTOC, denoted as $\eta$. If we follow the OTOC contours composed of same $\eta$ value, we obtain a series of space-time coordinates that give us a wavefront \cite{PhysRevB.96.020406,PhysRevLett.111.207202,PhysRevLett.111.260401,2018arXiv180506895L}. A couple of wavefronts associated with different $\eta$ values ranging between $\eta=1$ and $\eta=10^{-2}$ are shown in Fig. \ref{Fig5}. These wavefronts are expected to present how the correlations spread in the system over time. The outermost wavefront $\eta \sim 1$ corresponds to the lightcone, while $\eta \sim 0$ corresponds to butterfly cone in the literature \cite{2017JHEP...05..065M}. The wavefronts that we extracted follow power-law: $x \sim t^{\gamma}$ where $\gamma$ is dubbed as dynamical exponent. Fig. \ref{Fig7} shows a range of $\gamma$ changing from the low end of $\sim 0.5$ to the high end of $\sim 1.5$ with respect to $\eta$ for different system sizes. It is not clear if $\gamma$ would have a maximum in Fig. \ref{Fig7} due to the limitations in the data. We find a sublinear lightcone with $\gamma < 1$ where the spread is sub-ballistic. This observation aligns with the rare-region effects \cite{doi:10.1002/andp.201600326}. On the other hand, as the system scrambles, we observe that the wavefronts first become linear $\gamma=1$ and then passes to a superlinear region $\gamma > 1$ in Fig. \ref{Fig7}a. Therefore, the butterfly cones at $\eta \sim 0$ seem to differ significantly from the lightcone at $\eta \sim 1$. The wavefront structures that demonstrate the superlinear butterfly cones can be seen in Fig. \ref{Fig6a}. We plot the rates of the wavefronts in the inset of Fig. \ref{Fig7} where the sublinear lightcone ($\eta=0.99$) initially bounds the rest. Towards the scrambling time, the linear wavefront ($\eta=1$) seems to be the new bound on the wavefront rates. A range of sublinear wavefronts were detected in disordered Heisenberg chain before \cite{PhysRevB.96.020406}, implying the lightcone still differs from the butterfly cones in the dynamical exponent. Super-ballistic spread of correlations ($\gamma > 1$) has been previously observed in 1D spin chains with power-law decaying long-range interactions \cite{PhysRevLett.111.207202,PhysRevLett.111.260401,2018arXiv180506895L}. The ladder models can always be mapped to a path that passes through all the sites, e.g. zigzag or meander paths, so that 1D Jordan-Wigner transformation can be applied \cite{PhysRevB.69.104419}. Such mappings bring long-range interactions due to the Jordan-Wigner strings, which could explain the super-ballistic spread appearing later in time. We note that its rate remain insignificant compared to the faster wavefronts. It is an interesting direction to see if other ladder models present similar wavefront structures. Finally, we demonstrate the irregular wavefronts appearing in the spatial region \cite{2018arXiv180706086S} when the displacement is $\Delta x = 1-2$ in Fig. \ref{Fig6b}. The only fitted wavefronts are the sublinear wavefronts shown in Fig. \ref{Fig6b} as dotted-white lines, because the wavefronts start to exhibit irregularities later in time (solid-white lines). The irregularity appears between the origin and two sites away from it, as we observe that it takes significantly greater time for the information to spread $\Delta x = 2$ units compared to $\Delta x = 1$ unit in the time interval of $t\sim 0.5~[1/J_{||}]$ to $t \sim 2~[1/J_{||}]$. Hence it seems that the information spread slows down locally and temporarily (the jump feature in Fig. \ref{Fig6b}) before showing a sub-ballistic trend for $\Delta x > 2$. Furthermore, after $t \sim 2~[1/J_{||}]$ the jump feature is replaced by a constant line between $\Delta x = 1$ and $\Delta x = 2$ units, which points to a locally-scrambled region in the ladder while the information still spreads in the rest of the system at a finite rate. This unusual region-restricted scrambling continues until the whole ladder completely scrambles. Therefore, we conclude that different rare-region effects are at play in the ladder-XX, which calls for a more systematic future study.

\section{OTOC detection protocols}

The scrambling in the ladder-XX model can be detected via the interference measurement scheme on many-body states in optical lattices \cite{PhysRevLett.109.020505,islam2015measuring} or the interferometric measurement scheme \cite{PhysRevA.94.040302}. We detail both measurement schemes in the following subsections and elaborate on their advantages and disadvantages. Since both schemes need an experimental random initial state preparation, we first focus on the initial state preparation.

\subsection{Initial state preparation}
\begin{figure}
\centering
\subfloat[]{\label{Fig7c}\includegraphics[width=0.24\textwidth]{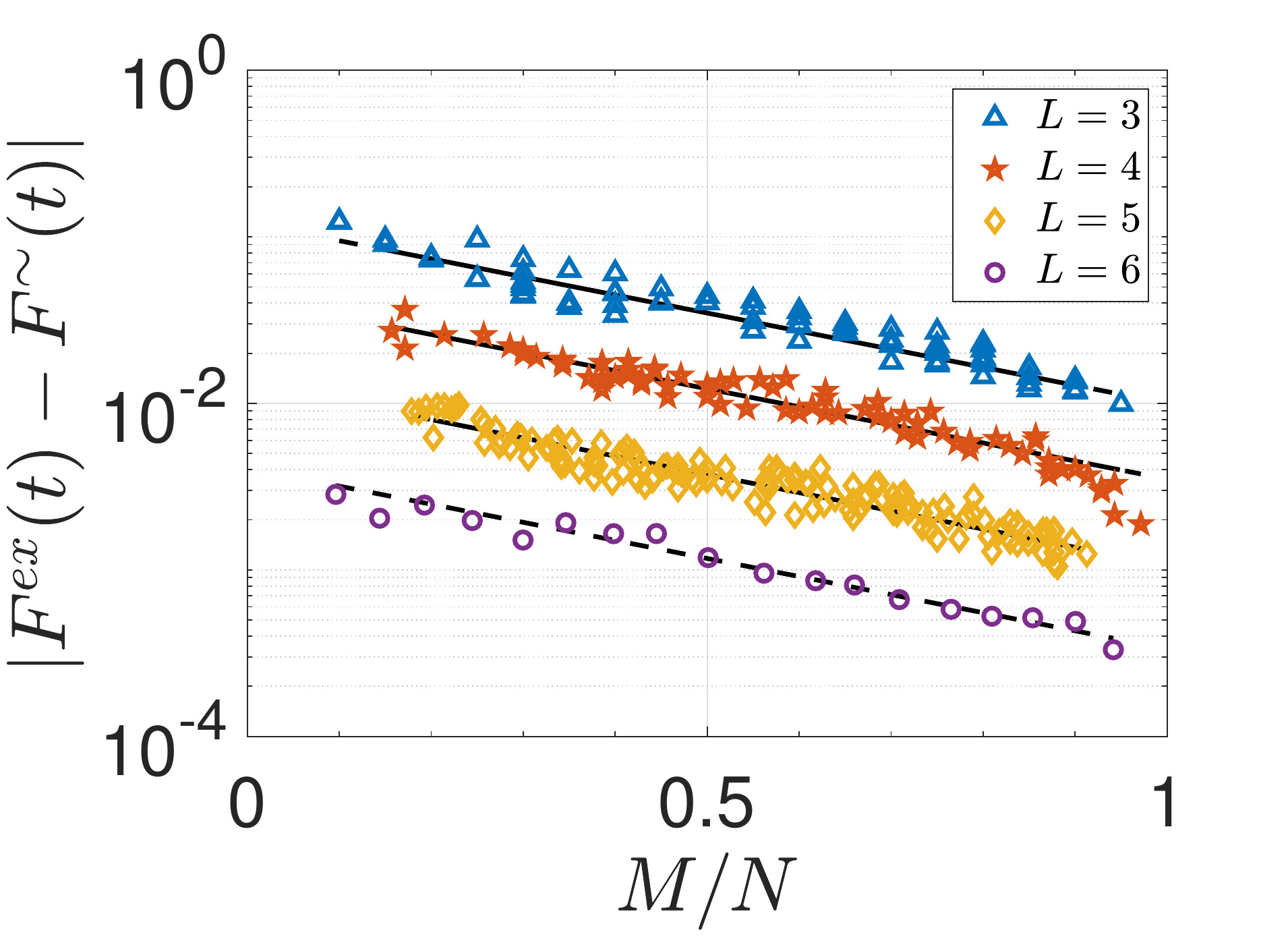}}\hfill 
\subfloat[]{\label{Fig7d}\includegraphics[width=0.24\textwidth]{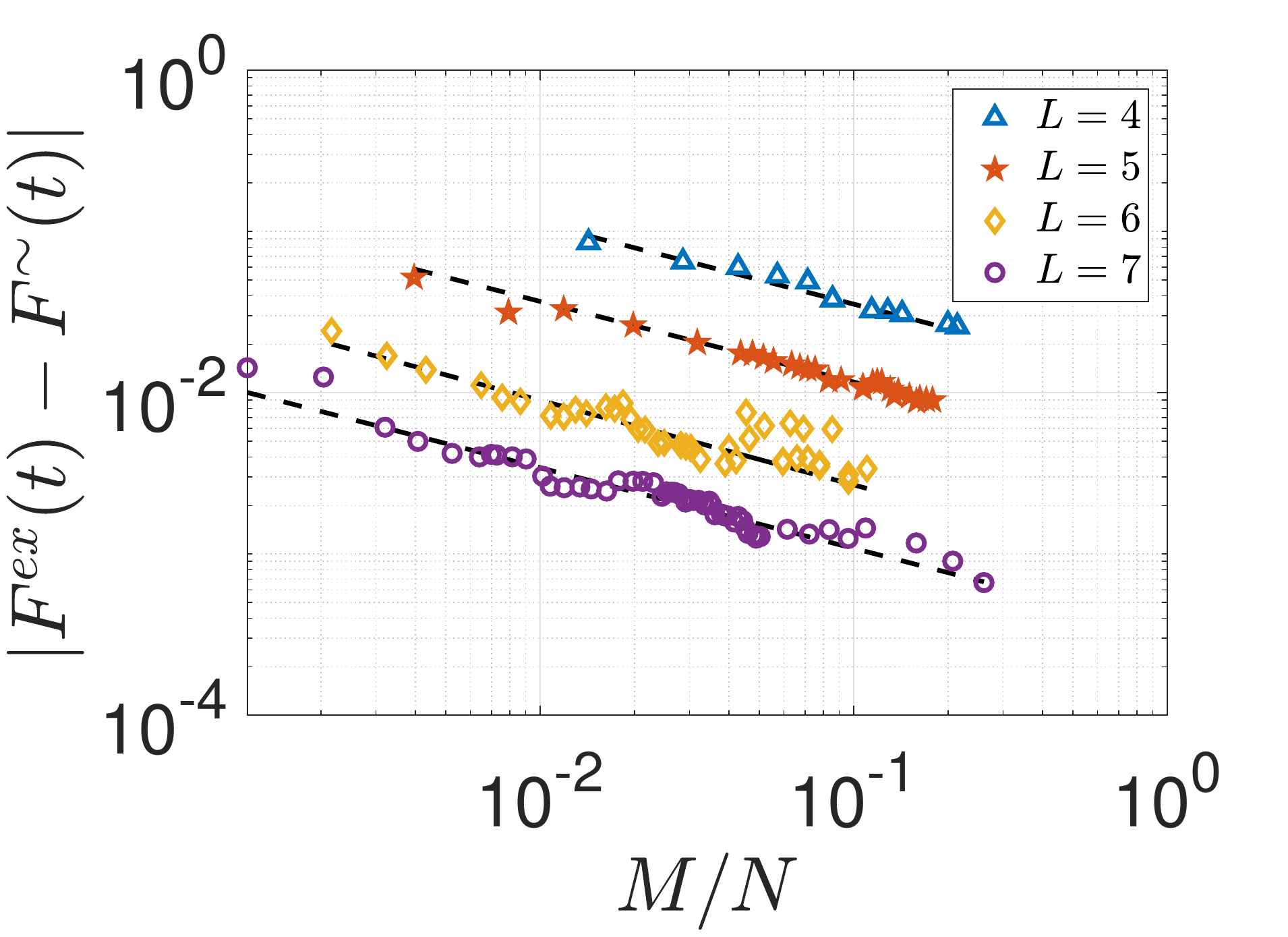}}\hfill 
\subfloat[]{\label{Fig7e}\includegraphics[width=0.24\textwidth]{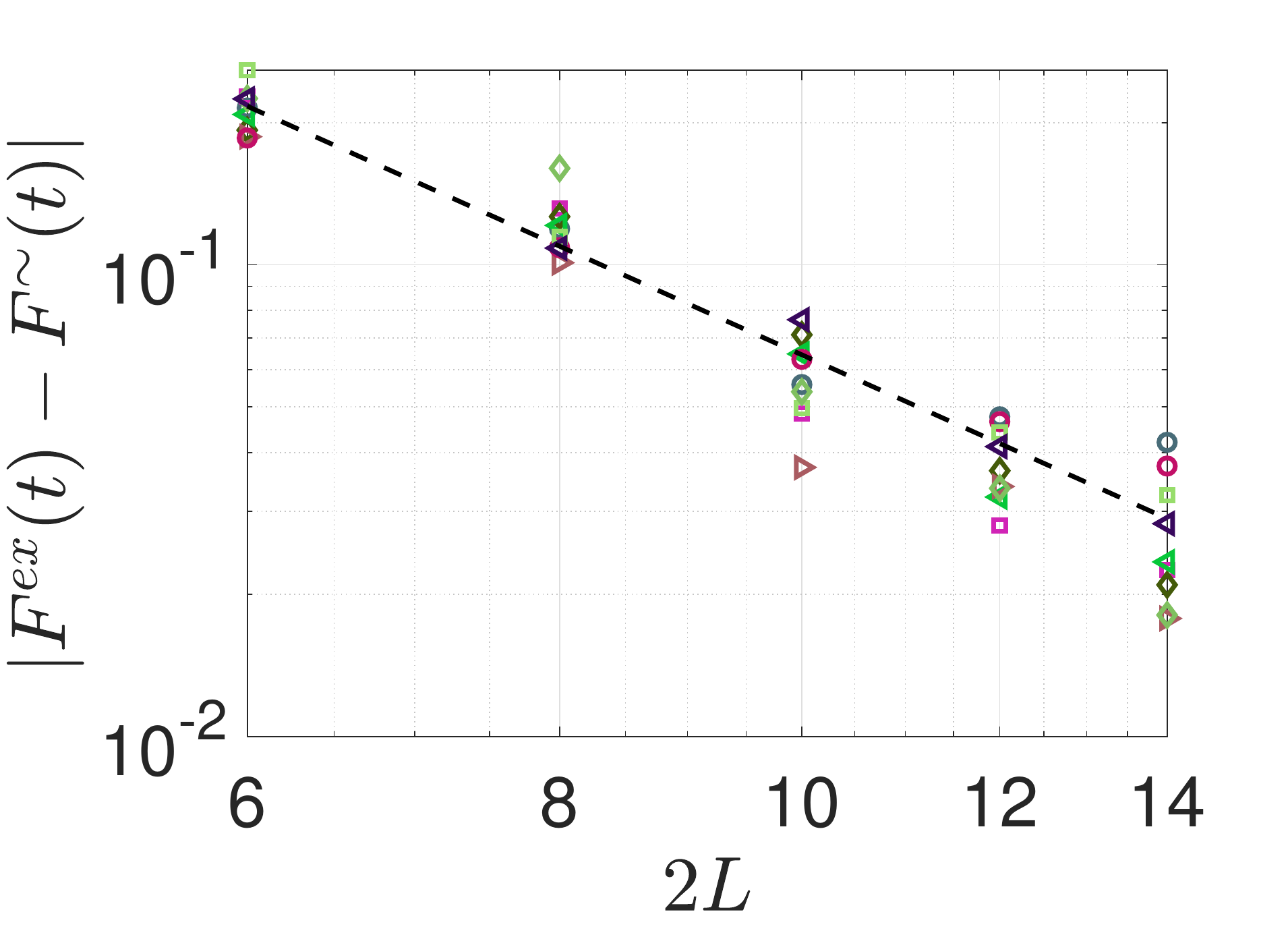}}\hfill 
\subfloat[]{\label{Fig7f}\includegraphics[width=0.24\textwidth]{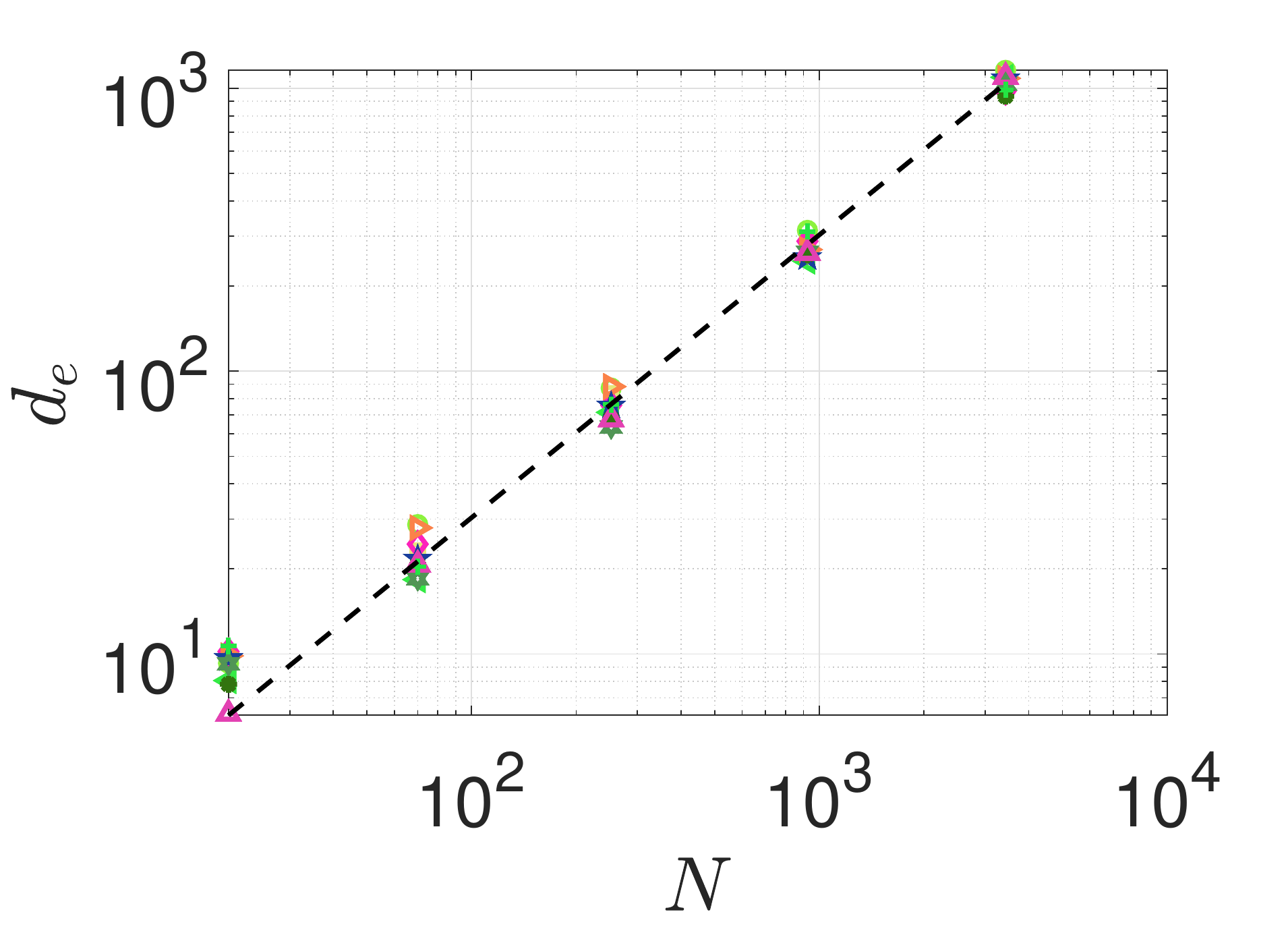}}\hfill 
\subfloat[]{\label{Fig7a}\includegraphics[width=0.24\textwidth]{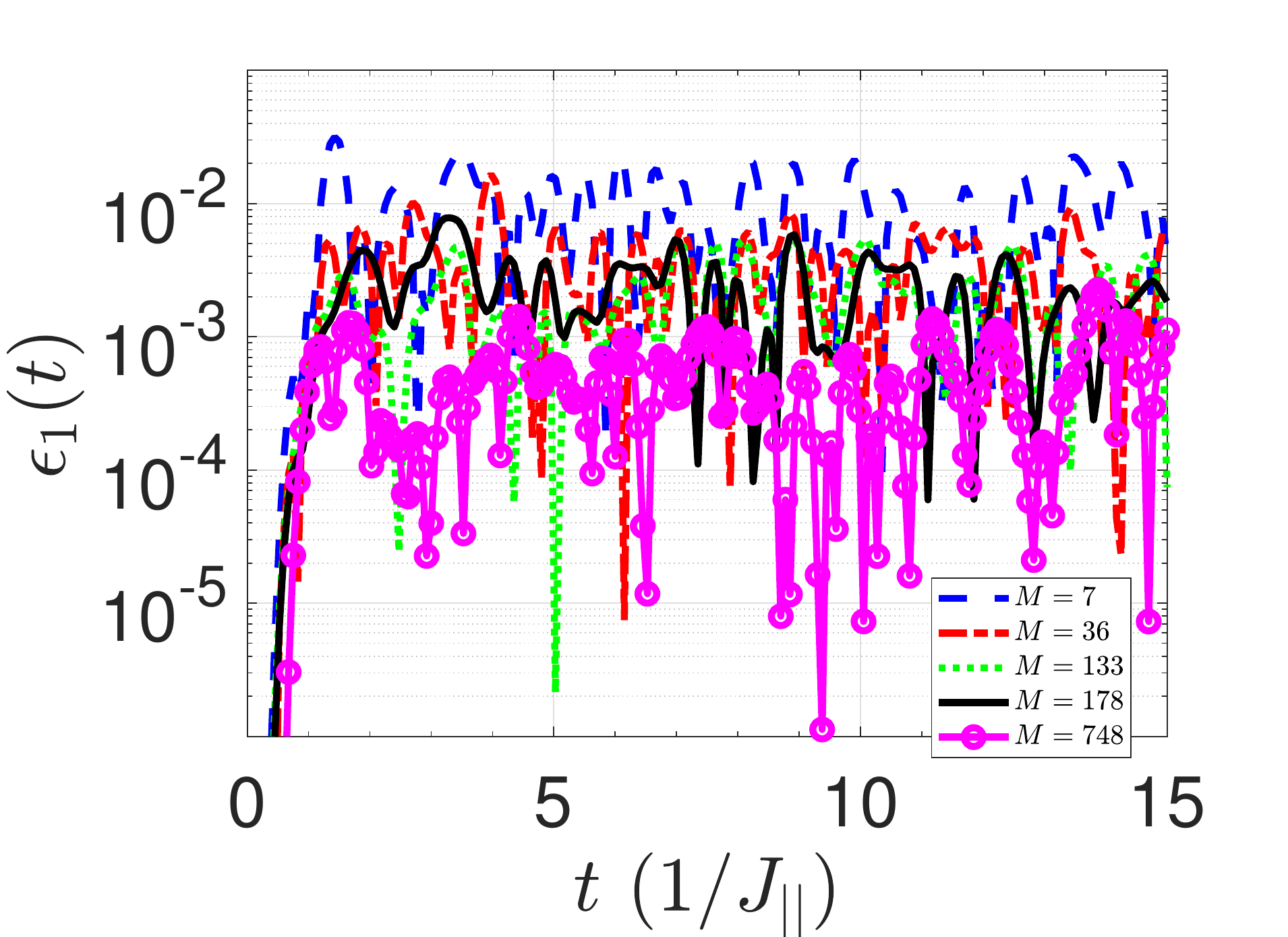}}\hfill 
\subfloat[]{\label{Fig7b}\includegraphics[width=0.24\textwidth]{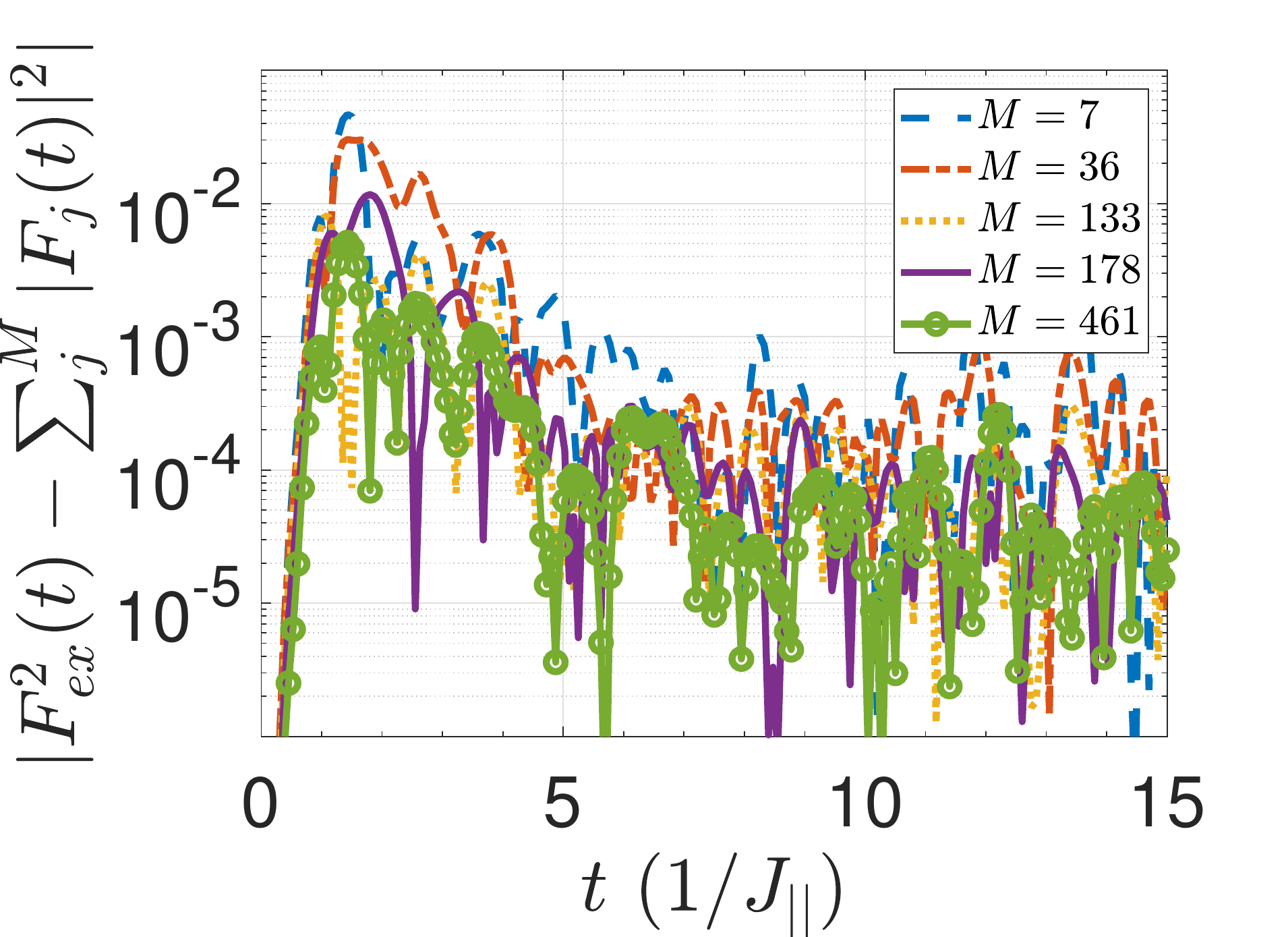}}
\caption{Initial state preparation at $h=1~[J_{||}]$. (a) The scaling of the mean error $\epsilon_1(t)$ with respect to $M/N$ sampling ratio, where $M$ and $N$ are the number of randomly-sampled states and the dimension of the Hilbert space, respectively. The blue-triangles, red-pentagrams, orange-diamonds and purple-circles stand for a single-chain size of $L=3$ to $L=6$, where all have an exponent of $b\sim-2.5$ in the fit $\propto a~\text{exp}\left(-b M/N\right)$. (b) The scaling of the mean error for small $M/N$ ratio has power-law scaling $\propto a\left(M/N\right)^b$ with $b\sim-0.5$ for all system sizes of $L=4$ (blue-triangles), $L=5$ (red-pentagrams), $L=6$ (orange-diamonds) and $L=7$ (purple-circles). (c)  The data collapse applied to the scaling of the mean of the error $\epsilon_1(t)$ with respect to the system size for only one randomly-sampled Fock state. Each data point is a random realization where the fitted curve gives an exponent of $b \sim - 2.26$ in $\epsilon_1(t) \propto (2L)^b$. (d) The scaling of the effective dimension, $d_e$ with the Hilbert space size, $N$, gives linear scaling $d_e = 0.3N$, mimicking an infinite-temperature state. (e) The error signal $\epsilon_1(t)$ with respect to time, for an average of $M=7$ (blue-dashed), $M=36$ (red-dashed dotted), $M=133$ (green-dotted) $M=178$  (black-solid) and $M=748$ (pink-circles) randomly-sampled Fock states. (f) The error signal $\epsilon_2(t)=||F^{\text{ex}}(t)|^2-\frac{1}{M}\sum_j^M|F_j(t)|^2|$ with respect to time, for an average of $M=7$ (blue-dashed), $M=36$ (red-dashed dotted), $M=133$ (orange-dotted) $M=178$  (purple-solid) and $M=461$ (green-circles) randomly-sampled Fock states. Both subfigures (e)-(f) have a system size of $L=6$.}
\label{Fig3}
\end{figure}
One can ideally use the whole set of Fock states to create a $\beta=0$ initial state. However, given this process would be lengthy, we ask if using a few $(M \ll N)$ randomly chosen Fock states would sufficiently mimic $\beta=0$ initial state $\mathbb{I} \sim \sum_{j=1}^M \Ket{\psi_j} \Bra{\psi_j}$, where $\Ket{\psi_j} = \left\lbrace\Ket{{{1 ...}\choose{0 ...}}},...,\Ket{{{0 ...}\choose{1 ...}}}\right\rbrace$ are Fock states for the ladder and they span the Hilbert space at half-filling. We find out that initiating an experiment with a randomly set Fock state for $\sim 10$ or $\sim 10^2$ times mimics the $\beta=0$ state up to a mean error of $\sim 7 \times 10^{-3}$ or $2 \times 10^{-3}$, respectively for a system size of $L=6$ (Fig. \ref{Fig7a}). We study how the mean error scales with the sampling ratio $M/N$ in Figs. \ref{Fig7c}-\ref{Fig7d} for different system sizes. Here the mean of the error is calculated for the data points when the error signal $\epsilon_1(t)=|F^{\text{ex}}(t)-\frac{1}{M}\sum_j F_j(t)|$ saturates in time. The sampling ratio $M/N$ has bounds $0<M/N<1$ and we observe when $M/N \gg 0$ the scaling is exponential and the data for all simulated system sizes could be collapsed to a single decay exponent $b\sim -2.5$ in $\epsilon_1(t) \propto a~\text{exp}\left(-b M/N\right)$, cf.~Fig.~\ref{Fig7c}. Note that when $M/N=1$, meaning that all Fock states are used, the error is zero up to machine precision and the OTOC is exact; and the point $M/N=0$ is not well-defined. Except for small sizes, e.g. $L=3$, the observed exponential scaling in Fig. \ref{Fig7c} is not experimentally practical due to the increasing number of randomly-sampled Fock states. Therefore, we study the limit $M/N \rightarrow 0$ separately where we obtain power-law scaling in $M/N$, cf. Fig.~\ref{Fig7d} with $b\sim -0.5$ in $\epsilon_1(t) \propto a\left(M/N\right)^b$ for system sizes $L=4-7$. 

Remarkably, it is possible to bound the error of approximation to $\sim 10^{-2}$ with only one Fock state for $L=7$. In fact the error decreases as a power-law with the increasing system size when only one Fock state is used to mimic the infinite temperature state (Fig.~\ref{Fig7e}). Fig.~\ref{Fig7e} shows 9 different realizations of using only one randomly-set Fock state and a single power-law curve fitted to all with $b \sim - 2.26$ in $\epsilon_1(t) \propto (2L)^b$ (App.~D). This observation is not utterly surprising, because a Fock state has a broad EON (eigenstate occupation number) distribution (Fig.~\ref{Fig7f} and App.~D). An EON distribution $|c_{\beta}|^2$ can be defined as the overlap of the initial state with the eigenbasis of the time-evolving Hamiltonian: $\Ket{\psi(0)}=\sum_{\beta}c_{\beta}\Ket{\psi_{\beta}} \rightarrow |c_{\beta}|^2$, where $\psi_{\beta}$ are the eigenstates and $\Ket{\psi(0)}$ is the initial Fock state. For instance, an infinite-temperature state has a uniform EON distribution: $|c_{\beta}|^2=1/N$. To be more precise, we can calculate the so-called effective dimension of the initial state, $d_e = \left(\sum_{\beta} |c_{\beta}|^4 \right)^{-1}$ \cite{2018PhRvA..97b3603D,Short_2011} and study the scaling of the effective dimension with the dimension of the Hilbert space. For an infinite-temperature state, $d_e = a N^{\xi}$ with an exponent of $\xi=1$ and $a=1$, which should be compared with the scaling exponent for the effective dimension of a randomly-set Fock state. Fig.~\ref{Fig7f} shows the data collapse on the effective dimensions of 10 different randomly-set Fock states for each system size. The fit parameters $d_e \sim 0.3N$ show that a randomly-set Fock state also gives an exponent $\xi=1$, which more accurately demonstrates the broadness of the EON distribution. The coefficient in front is bounded for effective dimension scalings, $a \leq 1$ and we see that a randomly-set Fock state has $a \sim 0.3$.  This reflects the fact that Fock state does not show uniform distribution in the eigenbasis of the Hamiltonian, and hence we have a nonzero error signal $\epsilon_1(t)$.

In conclusion, we see that the exact shape of the EON distribution is insignificant as $L \rightarrow \infty$, as long as it is a broad distribution in the eigenbasis. Therefore, only one Fock state could approximate the infinite-temperature OTOC reasonably well. We note that our analysis is valid for $h=1~[J_{||}]$ disorder strength. The observation that a single Fock state could exhibit $\xi=1$ exponent in its effective dimension scaling is possibly related to the extended eigenstates existing throughout spectrum in the chaotic regime. Hence, whether the found power-law scaling in system size for a single Fock state as well as the exponential and power-law scalings of the error in the sampling ratio $M/N$, depend on the disorder strength is an interesting question for future studies and experiments. Our results also show that a few randomly-sampled Fock states could be used as an alternative approach to Haar-distributed random states in numerics to calculate OTOC with a $\beta=0$ initial state at the chaotic regime of a model.

\subsection{The interference measurement}

\begin{figure}
\centering
\subfloat[]{\label{Fig4a}\includegraphics[width=0.49\textwidth]{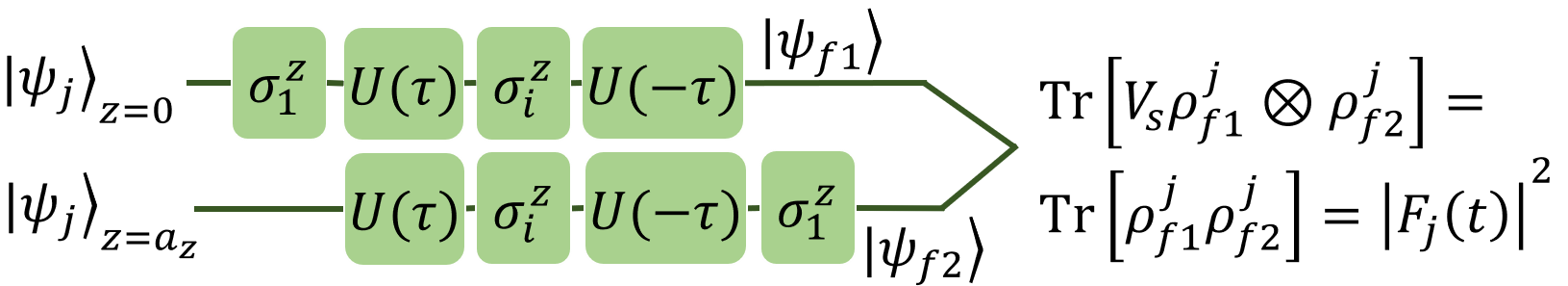}}\hfill \subfloat[]{\label{Fig4b}\includegraphics[width=0.40\textwidth]{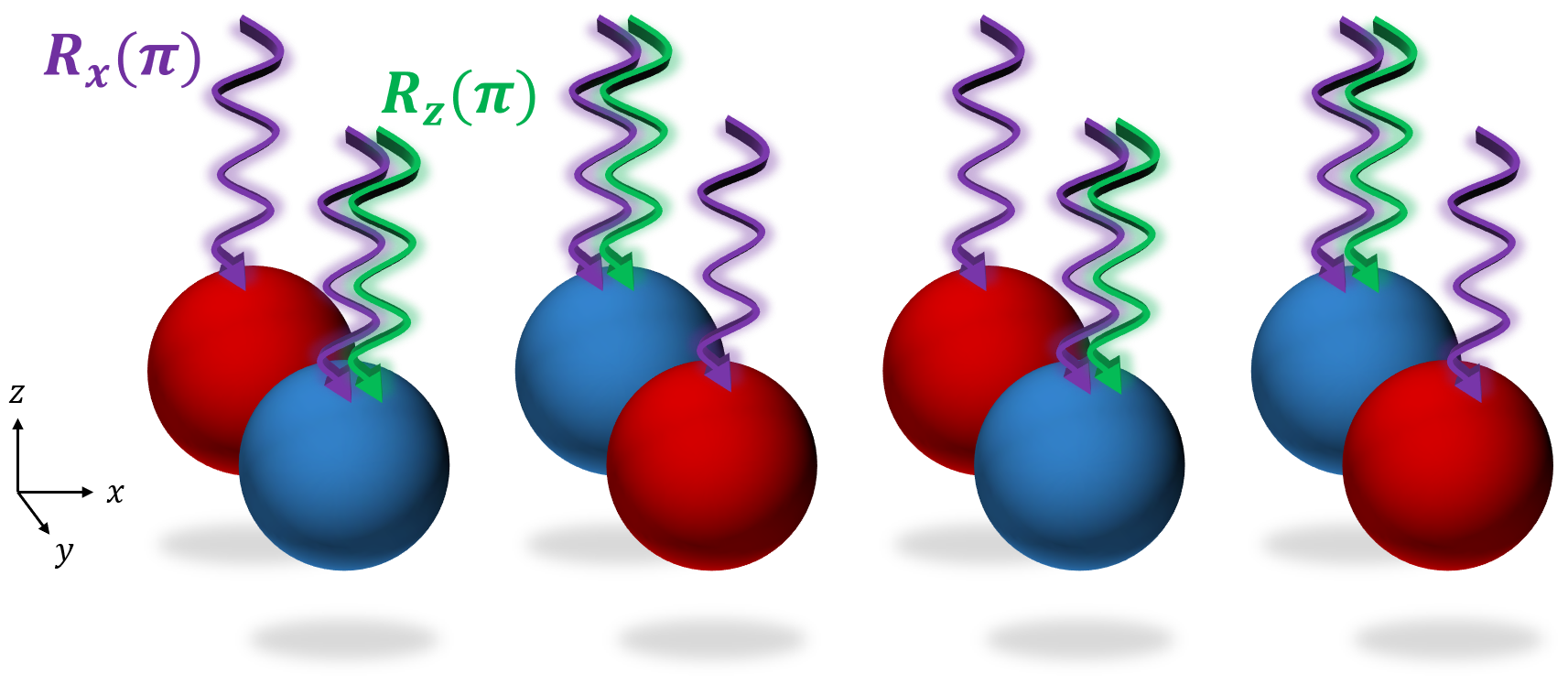}}
\caption{(a) The schematic that illustrates the circuit for OTOC measurement with the spin operators $\sigma_1^z$ and $\sigma_i^z$. The circuit utilizes interference measurements providing $\text{Tr}\left\lbrace \Ket{\psi_{f1}}\Braket{\psi_{f1}|\psi_{f2}}\Bra{\psi_{f2}} \right\rbrace = |F_j(\tau)|^2$. (b) Schematic for Hamiltonian sign-reversal protocol for evolution backwards in time: red and blue spheres stand for spin up and down states, respectively. We simultaneously perform $R_z(\pi) R_x(\pi)$ gates for the odd-numbered spins in the first leg and even-numbered spin in the second leg, while only one gate $R_x(\pi)$ is applied to the rest of the spins. $R_z(\pi)$ and $R_x(\pi)$ are denoted by green and purple wiggly lines, meaning that the single-spin gates for cold-atom systems could be realized via laser pulses \cite{2006PhRvA..74d2316Z,2014NatSR...4E5867L} or microwaves \cite{2013NatCo...4E2027L}.}
\label{Fig4}
\end{figure}

$|F(t)|^2$ is the quantity to measure in the interference measurement scheme \cite{PhysRevLett.109.020505}. We see that $\text{Im}(F(t)) \sim 0$ and $\text{Re}(F(t)) \geq 0$ throughout the simulation time with the parameters used in the paper, thus rendering $|F(t)|^2$ a good quantity to measure. The interference measurement scheme has been proposed to probe scrambling in Bose-Hubbard model previously \cite{2017PhRvB..96e4503S,2017NJPh...19f3001B}, however note that the implementation of the interference measurement further simplifies for the hard-core boson limit \cite{PhysRevLett.109.020505} which we utilize in the cold-atom setup of our model. The steps of the interference detection protocol follow as (Fig. \ref{Fig4a}):

(i) Generate two copies of the same randomly-sampled Fock state $\Ket{\psi_j}$: We can first set a 2D lattice to Mott-insulator phase with unit filling factor and then adiabatically ramp the lattice potential to a double-well potential at each site in the y-direction. This would produce $\left(\Ket{10}+\Ket{01}\right)/\sqrt{2}$ state for a double-well; and via suppressing the tunneling between wells in the double-wells, one can generate randomly sampled Fock states in 2D lattice at half-filling. To make two copies of the initial state, we can introduce another lattice layer in z-direction and apply the same operations of lattice potential simultaneously for both planes.

(ii) Apply $\sigma^z_{1,1}$ gate on the first spin in the lower leg in the first copy. 

(iii) Apply to both copies $U(\tau) \sigma^z_{1,i}$, where $U(\tau)$ is evolution forward in time for $\tau$ and $\sigma^z_i$ gate is applied to any spin $i$ further away from the first spin in the lower leg.

(iv) \emph{Hamiltonian sign reversal protocol}: As illustrated in Fig. \ref{Fig4b}, we apply a set of gates to the lattice sites simultaneously to change the overall sign of the Hamiltonian so that we could evolve the many-body state with $-H$. Given that we shine either laser pulses \cite{2006PhRvA..74d2316Z,2014NatSR...4E5867L} or microwaves \cite{2013NatCo...4E2027L} to implement single-spin rotations, our protocol of Hamiltonian sign-reversal could be related to NMR (nuclear magnetic resonance) Hamiltonian engineering \cite{PhysRevX.7.031011,PhysRevLett.120.070501}, though with a difference of site-resolving pulses in the cold-atom setup. Remembering $R_z^{\dagger}(\theta) \sigma^x R_z(\theta) \rightarrow \cos \theta \sigma^x - \sin \theta \sigma^y$, $R_z^{\dagger}(\theta) \sigma^y R_z(\theta) \rightarrow \cos \theta \sigma^y + \sin \theta \sigma^x$, we can create sign difference in the X and Y coupling terms if we apply the $R_z(\pi)$ pulse alternating on the sites, e.g. odd-numbered and even numbered spins in the first and second legs, respectively. In order to change the sign of the random disorder term, we apply $R_x(\pi)$ gate to each of the spins via utilizing $R_x^{\dagger}(\theta) \sigma^y R_x(\theta) \rightarrow \cos \theta \sigma^y - \sin \theta \sigma^z$. Then the gate sequence that we apply to both copies becomes, 
\begin{eqnarray}
\Pi_{i:odd} R^z_{1,i} R^x_{1,i} R^x_{1,i+1} R^z_{2,i+1} R^x_{2,i} R^x_{2,i+1}(\pi),
\label{signReversal}
\end{eqnarray}
where $1-2$ denotes the leg number. Eq. \ref{signReversal} could be realized via a programmable acousto-optic modulator (AOM) with multiple laser outputs whose frequency differences are negligible \cite{Aldous:17} and high-resolution imaging devices that can provide single-site addressability \cite{2009Natur.462...74B,2013NatCo...4E2027L}.

(v) Apply $\sigma^z_{1,1}$ gate on the first spin in the second copy. 

(vi) Make an interference measurement between final copies $\Ket{\psi_{f1}}=U(-\tau)\sigma^z_{1,i}U(\tau)\sigma^z_{1,1}\Ket{\psi_j}$ and $\Ket{\psi_{f2}}=\sigma^z_{1,1} U(-\tau)\sigma^z_{1,i}U(\tau)\Ket{\psi_j}$ in the hard-core boson limit \cite{PhysRevLett.109.020505,islam2015measuring}. By measuring the swap operator on both copies \cite{PhysRevLett.109.020505}, we can obtain $\text{Tr}\left\lbrace \rho_{f1}\rho_{f2} \right\rbrace = |F_j(\tau)|^2$ for each $\Ket{\psi_j}$ initial state where $\rho_{f1}=\Ket{\psi_{f1}}\Bra{\psi_{f1}}$. The same measurement could be applied to the copies of initial state to check if they are identical $\text{Tr}\left\lbrace \rho_j^2 \right\rbrace = 1$. The interference measurement scheme has been applied to measure entanglement entropy \cite{islam2015measuring}.

(vii) Repeat the measurement protocol for $M$ times with randomly chosen $\Ket{\psi_j}$ initial states to obtain $\frac{1}{M}\sum_j|F_j(\tau)|^2$ which is equal to $|F^{\text{ex}}(t)|^2$ up to an error $\lesssim 10^{-2}$ and $\sim 10^{-4}$ in decay and saturation regimes, respectively for $M\sim 10^2$ Fock states. Fig. \ref{Fig7b} shows the difference between the square of the exact OTOC (Eq. \ref{OTOC}) and $\frac{1}{M}\sum_j^M|F_j(\tau)|^2$ for M randomly chosen Fock states for a system size $L=6$.

\subsection{The interferometric scheme}

\begin{figure}
\centerline{\includegraphics[width=0.49\textwidth]{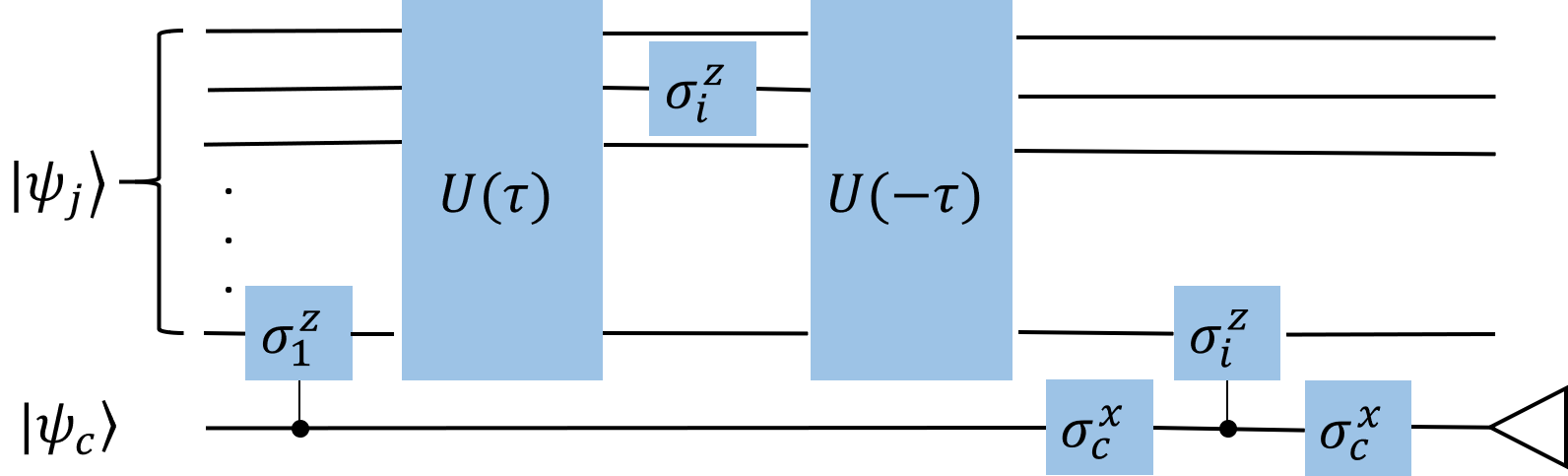}}
\caption{The measurement circuit for the interferometric approach \cite{PhysRevA.94.040302} on the ladder-XX model with local spin observables $\sigma_1^z$ and $\sigma_i^z$ by using an auxiliary spin $\Ket{\psi_c}$ to measure only the real part of the OTOC.}
\label{Fig8}
\end{figure}

We can measure $F(t)$ with the interferometric approach  \cite{PhysRevA.94.040302}, because the measurement of the control spin either in x- or y-basis provides the real and imaginary parts of the OTOC, respectively. Fig. \ref{Fig8} demonstrates the measurement circuit where the control spin needs to be coupled only to the first spin in the ladder. The protocol follows as:

(i) Initialize the control spin in a superposition state of $\Ket{\psi_c} = \left(\Ket{0}_c + \Ket{1}_c\right)/\sqrt{2}$ to prepare the many-body state
\begin{eqnarray}
\frac{1}{\sqrt{2}}\left[ \left(\sigma^z_{1,1} \sigma^z_{1,i} (t) \Ket{\psi_j} \right) \Ket{0}_c +  \left( \sigma^z_{1,i}(t) \sigma^z_{1,1} \Ket{\psi_j} \right) \Ket{1}_c \right], \notag
\end{eqnarray}
where the ladder-XX model is simultaneously initiated in a randomly-sampled Fock state $\Ket{\psi_j}$.

(ii) Apply controlled-$\sigma^z_1$ operation to the first spin in the lower leg: $\left(\Ket{0}_c\Bra{0} \otimes \mathbb{I}_1 + \Ket{1}_c\Bra{1} \otimes R^z_{1,1}(\pi)\right) \otimes \mathbb{I}^{\otimes 2L-1}$.  

(iii) Evolve the ladder-XX model forward in time and apply $\sigma^z_i$ rotation to the spin $i$: $ \mathbb{I}_c\otimes U(\tau) \left(\mathbb{I}^{\otimes i-1} \otimes \sigma^z_{1,i} \otimes \mathbb{I}^{\otimes 2L-i}\right)$.

(iv) Apply Eq. \ref{signReversal} to the ladder-XX model and evolve the many-body state with $-H$ as $ \mathbb{I}_c\otimes U(-\tau)$.

(v) Apply $\sigma^x_c$ gate to the control spin before another controlled-$\sigma^z_1$ operation, so that we have $\left(\Ket{0}\Bra{0}_c \otimes R^z_{1,1}(\pi) + \Ket{1}\Bra{1}_c \otimes \mathbb{I}_1  \right) \otimes \mathbb{I}^{\otimes 2L-1}$. Further apply another $\sigma^x_c$ gate to the control spin.

(vi) Make a measurement on the control spin in the x-basis to obtain the real part of the OTOC, $\text{Re}\left[F_j(t)\right] = \Braket{\sigma^x_c} = \Bra{\psi_j(t)} \sigma^x_c \Ket{\psi_j(t)}$. 

(vii) Repeat the measurement protocol for $M$ times with randomly chosen $\Ket{\psi_j}$ initial states to obtain $\frac{1}{M}\sum_j F_j(\tau)$ which is equal to $F^{\text{ex}}(t)$ up to an error shown in Fig. \ref{Fig3}.

\emph{Outlook}. The interference measurement scheme requires two copies of the same randomly-sampled initial Fock state, which is challenging but doable. On the other hand, the interferometric approach could be realized with only one copy. However, in this measurement scheme we need to couple an auxiliary spin to the first spin and implement controlled-spin gates \cite{PhysRevLett.104.010503,2015Natur.527..208K} which is challenging in the current technology. Therefore both approaches have certain (dis-)advantages. An important difference that we observe in two measurement schemes are the error bounds due to the measurement output, $|F^{\text{ex}}(t)|^2=\frac{1}{M}\sum_j^M|F_j(t)|^2$ and $F^{\text{ex}}(t)=\frac{1}{M}\sum_j F_j(t)$ for interference and interferometric, respectively. The error bounds are stable throughout the evolution in the interferometric approach; while they significantly lower in the saturation regime (by a factor of $\sim 10^2$) and slightly higher in the decay regime of an interference measurement. Therefore, in the case of measuring only the saturation values of the OTOC, the interference measurement seems to be more advantageous.

\section{Conclusions}

The ladder-XX model's OTOC decay profiles and information spread show a variety of phenomena ranging from quantum chaos to MBL phase and possibly rare-region effects in the ergodic phase that we leave as a future study. We further discussed a Hamiltonian sign reversal protocol that is a novel alternative to existing approaches in cold atoms and how to apply both interference and interferometric measurements in the scrambling detection with experimental random state preparation. Our results demonstrate that the experiments could utilize only one randomly-set Fock state for sufficiently big many-body systems to reproduce infinite-temperature OTOC up to a bounded error in the chaotic regime. The XX-ladder has a more convenient experimental cold-atom setup compared to the Heisenberg chain, since it lacks Z-coupling terms, while it is still interacting due to its quasi-1D nature. Thus, it can be more easily implemented in the laboratory to further investigate scrambling and understand how scrambling changes in the transition from 1D to 2D.

\section{Acknowledgements} 
This work was supported by the AFOSR MURI program. C.B.D. thanks P. Myles Eugenio for interesting discussions and comments on the manuscript and on invariant subspaces of the ladder-XX models; Minh Tran for helpful discussions on superlinear lightcones; Zheng-Hang Sun for helpful discussions on MBL phase in the ladder models.

\appendix

\section{Error bars on OTOC for the disordered XX-ladder}

\begin{figure}
\centerline{\includegraphics[width=0.45\textwidth]{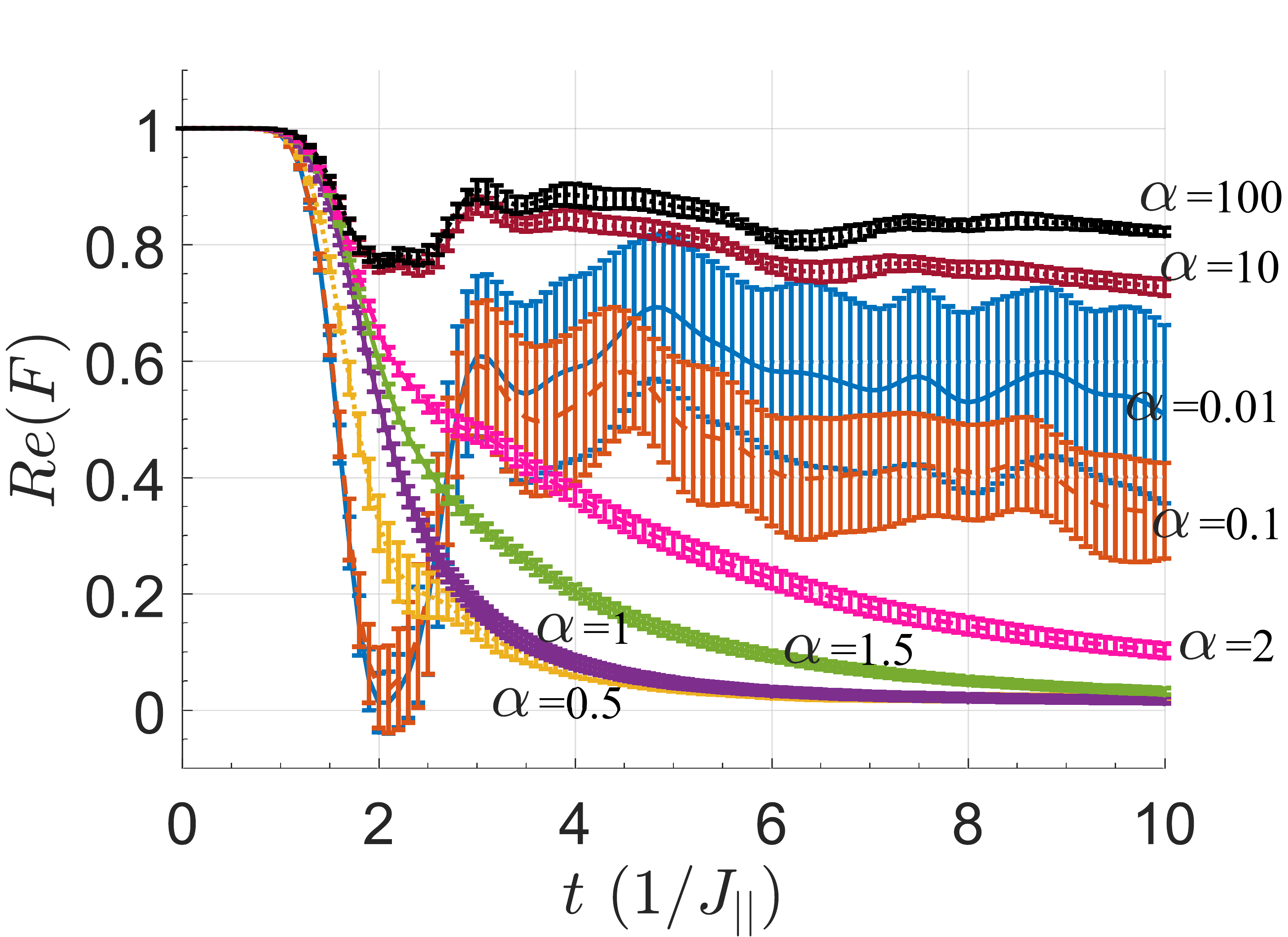}}
\caption{Error bars of the out-of-time order correlators with disorder strength of $h = 1~[J_{||}]$ between two distant operators $\sigma^z_1$ and $\sigma^z_7$ with respect to different rung interaction strengths $\alpha$ where $J_{\perp} = \alpha J_{\parallel}$ for $L=7$. The OTOC is averaged over 100 different random samples. The curves are, $\alpha=0.01$ (blue-solid), $\alpha=0.1$ (orange-dashed), $\alpha=0.5$ (yellow-dotted), $\alpha=1$ (purple-solid), $\alpha=1.5$ (green-solid), $\alpha=2$ (pink-dashed), $\alpha=10$ (crimson-dotted) and $\alpha=100$ (black-dotted).}
\label{SuppFig6}
\end{figure}

Fig. \ref{SuppFig6} shows the out-of-time-order correlators for different rung couplings with error bars in the case of $h=1~[J_{||}]$ random disorder strength. The error bars are significant for smaller rung couplings where the integrable limit of the ladder-XX model resides. 
As the rung coupling becomes equal to intra-leg couplings, the error bars become smaller. Therefore, the scrambling that we observe in the chaotic limit is robust to different configurations with the random disorder strength of $h\sim 1~[J_{||}]$. The error bars are more pronounced in the decay compared to unity and saturation regimes. When we study the opposite regime of dimer phase where rung coupling is much bigger than the intra-leg coupling $\alpha \rightarrow \infty$, the error bars do not grow significantly. 

\section{Error bounds on Haar-distributed initial states}

\begin{figure}
\centerline{\includegraphics[width=0.45\textwidth]{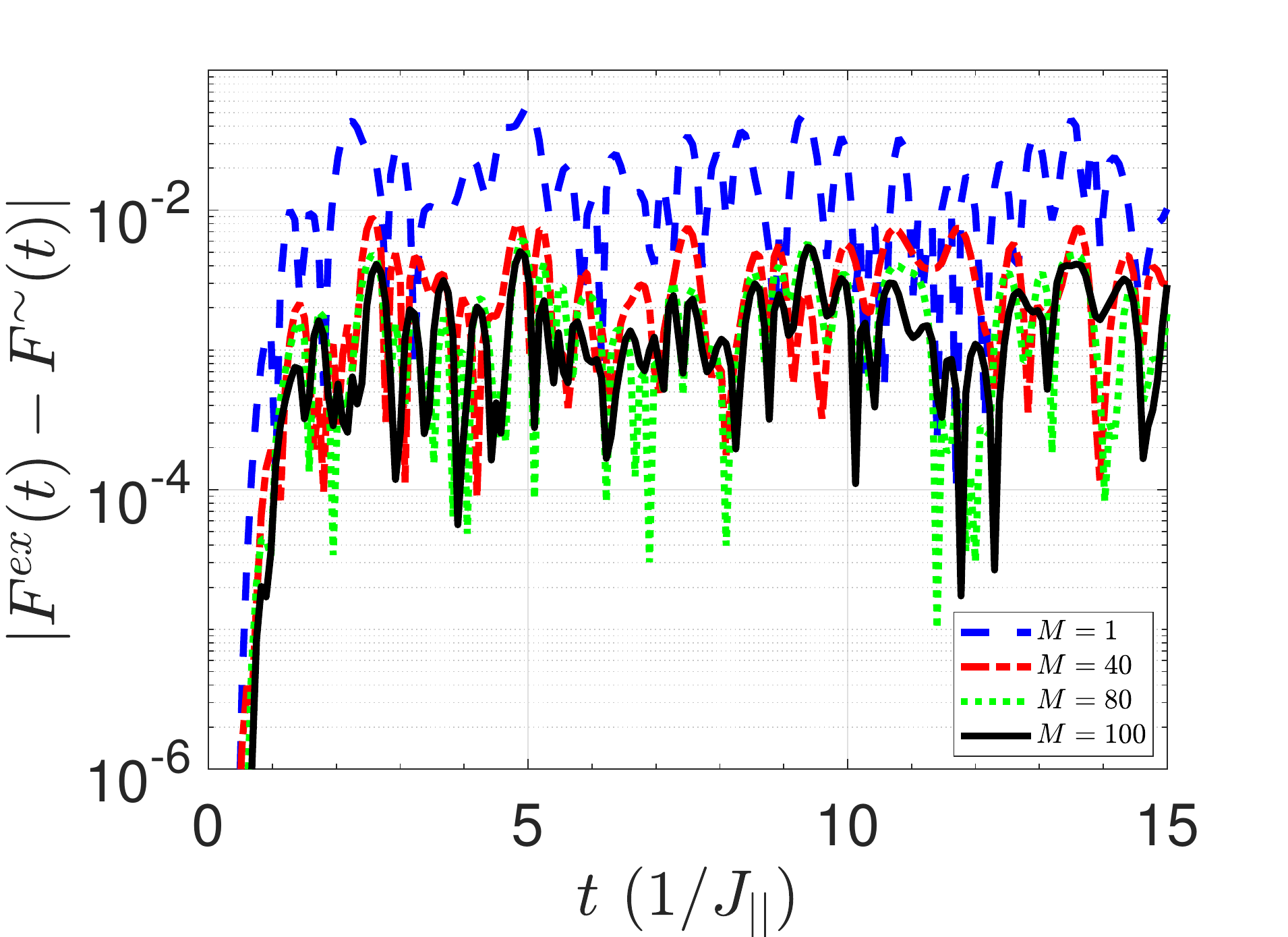}}
\caption{The difference $|F_{i=6}^{\text{ex}}(t)-F_{i=6}^{\sim}(t)|$ for only one Haar-distributed random state (blue-dashed), averaged over 40 random states (red-dashed dotted), 80 states (green-dotted) and 100 states (black-solid). Only the real part of $F_{i}^{\sim}(t)$ is taken since the imaginary part is practically zero.}
\label{SuppFig11}
\end{figure}

We present the error bounds on the OTOC when Haar random states are used to mimic the $\beta=0$ initial state in Fig. \ref{SuppFig11}. Fig. \ref{SuppFig11} shows the difference $|F_i^{\text{ex}}(t)-F_{i}^{\sim}(t)|$ for $L=6$ system size at $h=1$ random disorder strength with only one random field configuration when $i=6$ is set. The blue line stands for the case where we take only one random initial state, whereas the black line shows the case where we average over 100 such initial states. The difference is slightly more than an order of magnitude. However as seen from the other curves, the mixture of a couple of them is quite close to the case with $M=100$. While using only one random state approximates the OTOC with an error up to $10^{-2}$, one can improve the error bound via averaging over only a few states. The results are obtained in this paper with an average of 100 random states.

\section{The exponential and power-law fitting parameters}

\begin{figure}[H]
\centering
\subfloat[]{\label{SuppFig3}\includegraphics[width=0.24\textwidth]{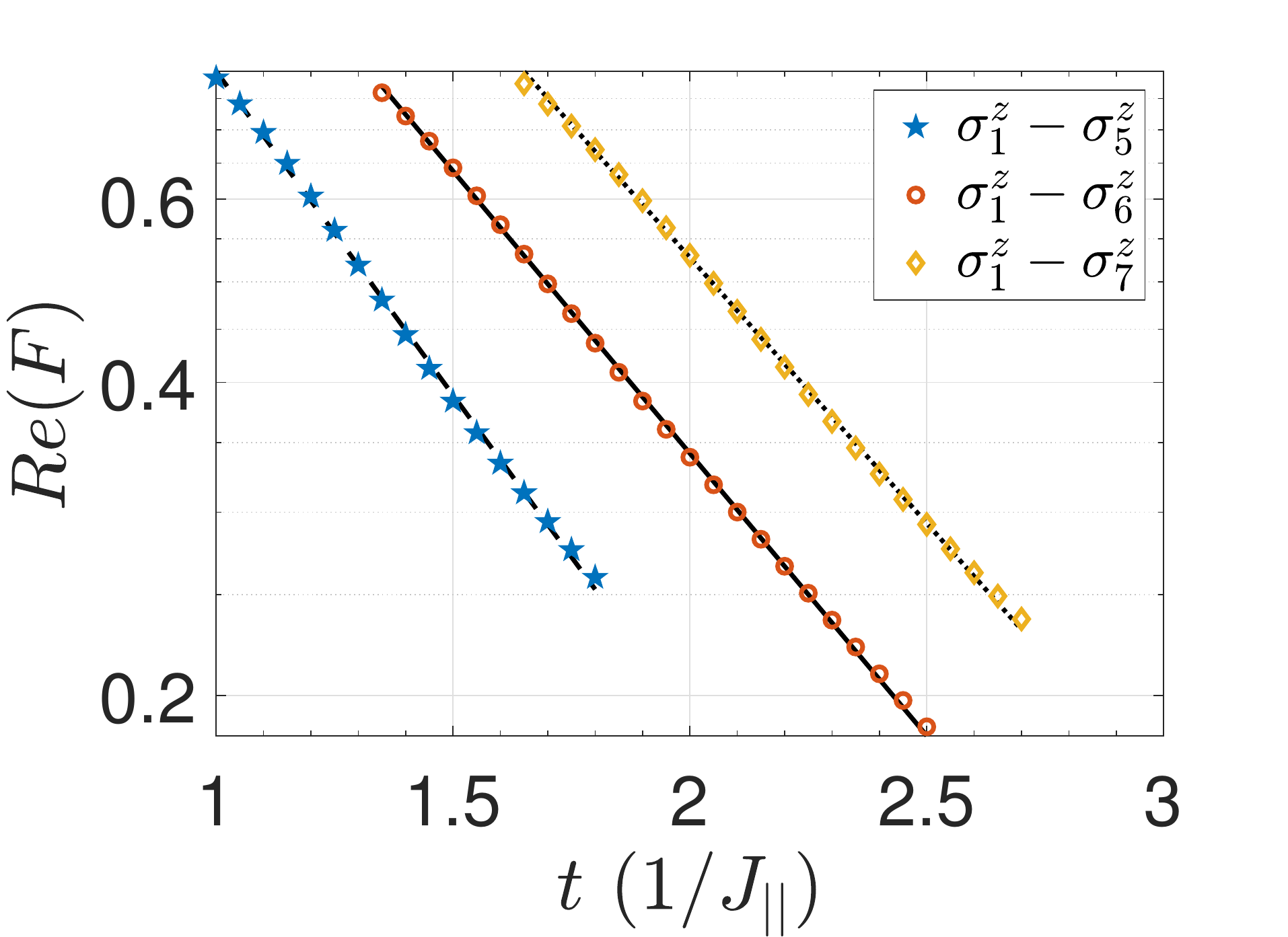}}\hfill \subfloat[]{\label{SuppFig4}\includegraphics[width=0.24\textwidth]{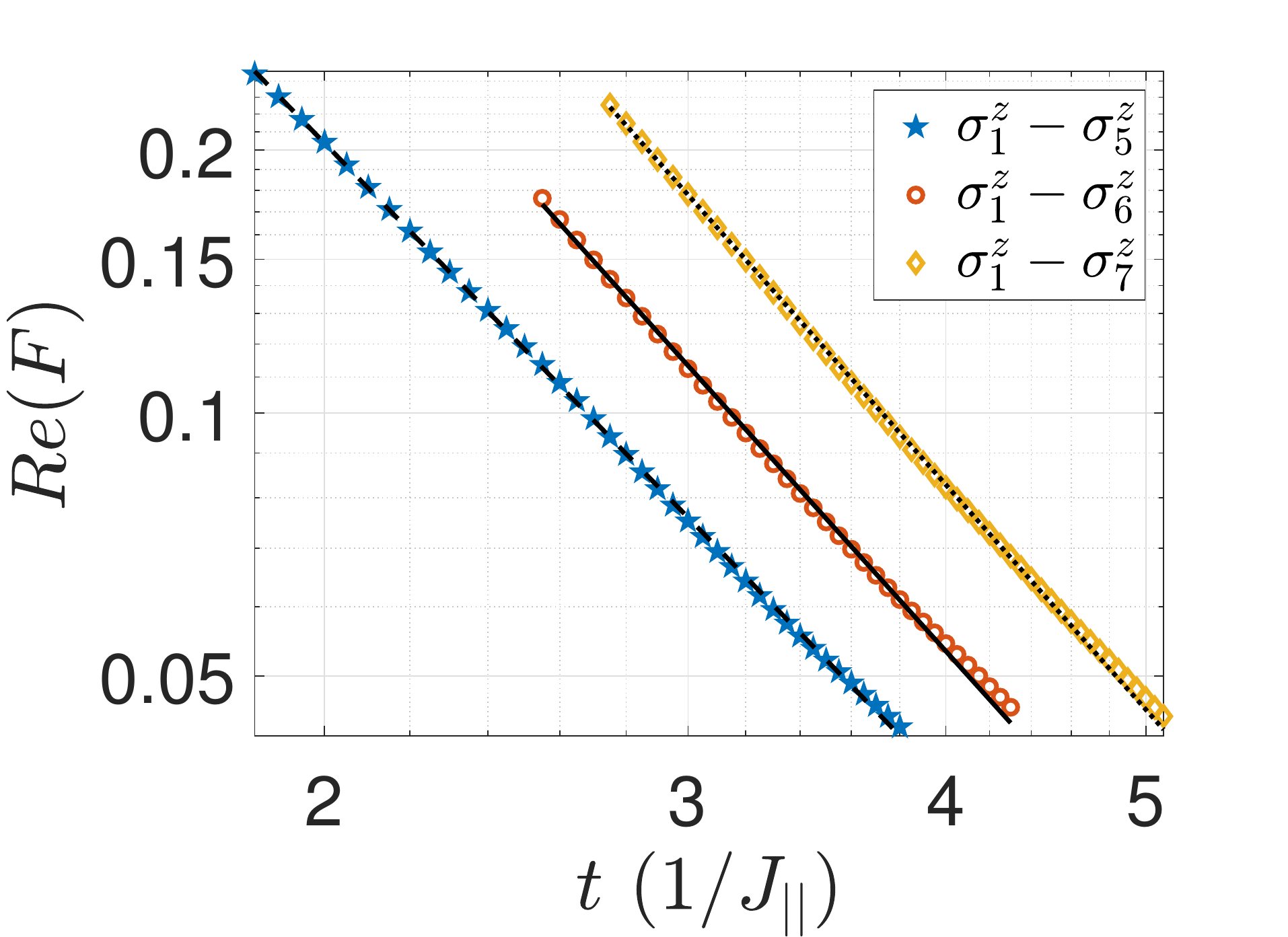}}\hfill 
\caption{(a) Semi-logarithmic plot for $\sigma_1^z$ with $\sigma_5^z$ (blue-pentagrams), $\sigma_6^z$ (red-circles) and $\sigma_7^z$ (orange-diamonds) observables in a system size of $L=7$.  The lyapunov-like exponents follow as, 1.4342 ($R^2 =0.9989$), 1.2507 ($R^2= 0.9996$) and 1.1767 ($R^2= 0.9994$) for $\sigma_5^z$-$\sigma_7^z$ with dashed, solid and dotted lines respectively. (b) Logarithmic plot for $\sigma_1^z$ with $\sigma_5^z$ (blue-pentagrams), $\sigma_6^z$ (red-circles) and $\sigma_7^z$ (orange-diamonds) observables in a system size of $L=7$. The power-law exponents follow as 2.4335 ($R^2 =0.9999$), 2.6165 ($R^2=0.9991$) and  2.6565 ($R^2=0.9997$) for $\sigma_5^z$-$\sigma_7^z$ with dashed, solid and dotted lines respectively. The data is averaged over 100 different realizations of the Hamiltonian at $h=1~[J_{||}]$ for both subfigures.}
\end{figure}

Here we present the additional figures and fitting data that show the exponential and power-law decays. Figs. \ref{SuppFig3}-\ref{SuppFig4} are for $L=7$ system size. 
The lyapunov-like exponents for $L=8$ follow as, 1.362 ($R^2 =0.9986$), 1.229 ($R^2=0.9992$), 1.09 ($R^2=0.9997$), 1.015 ($R^2=0.9996$) for $\sigma_5^z$-$\sigma_8^z$ respectively (the figure is shown in the main text). The power-law exponents follow as 2.1865 ($R^2 =0.9996$), 2.5506 ($R^2=0.9981$),  2.5751 ($R^2=0.9976$), 2.7636 ($R^2=0.9995$) for $\sigma_5^z$-$\sigma_8^z$ respectively. The data is averaged over 10 different random samples all at $h=1$. We also note that the interval of data used for exponential fitting when $L=8$ is from the time when OTOC starts to deviate from unity through $t \sim 2~[1/J_{||}]$, $t \sim 3$ [1/J], $t \sim 4~[1/J_{||}]$ and $t \sim 4~[1/J_{||}]$ for $\sigma^z_5$, $\sigma^z_6$, $\sigma^z_7$ and $\sigma^z_8$, respectively. The power-law fitting is applied to data seen in Fig. 2b (in main text) until $t\sim 5~[1/J_{||}]$, $t\sim 5~[1/J_{||}]$, $t\sim 6~[1/J_{||}]$ and $t\sim 6~[1/J_{||}]$ for $\sigma^z_5$, $\sigma^z_6$, $\sigma^z_7$ and $\sigma^z_8$, respectively. Similarly, the data used for the power-law in the clean limit, $h=0$, is shown in Fig. 2c in the main text (until $t\sim 10~[1/J_{||}]$ for all operators). The MBL decay form is applied to all data as seen in Fig. 2d in the main text.

\section{Details on the experimental initial state preparation}

\begin{figure}
\centering
\subfloat[]{\label{supFig1}\includegraphics[width=0.24\textwidth]{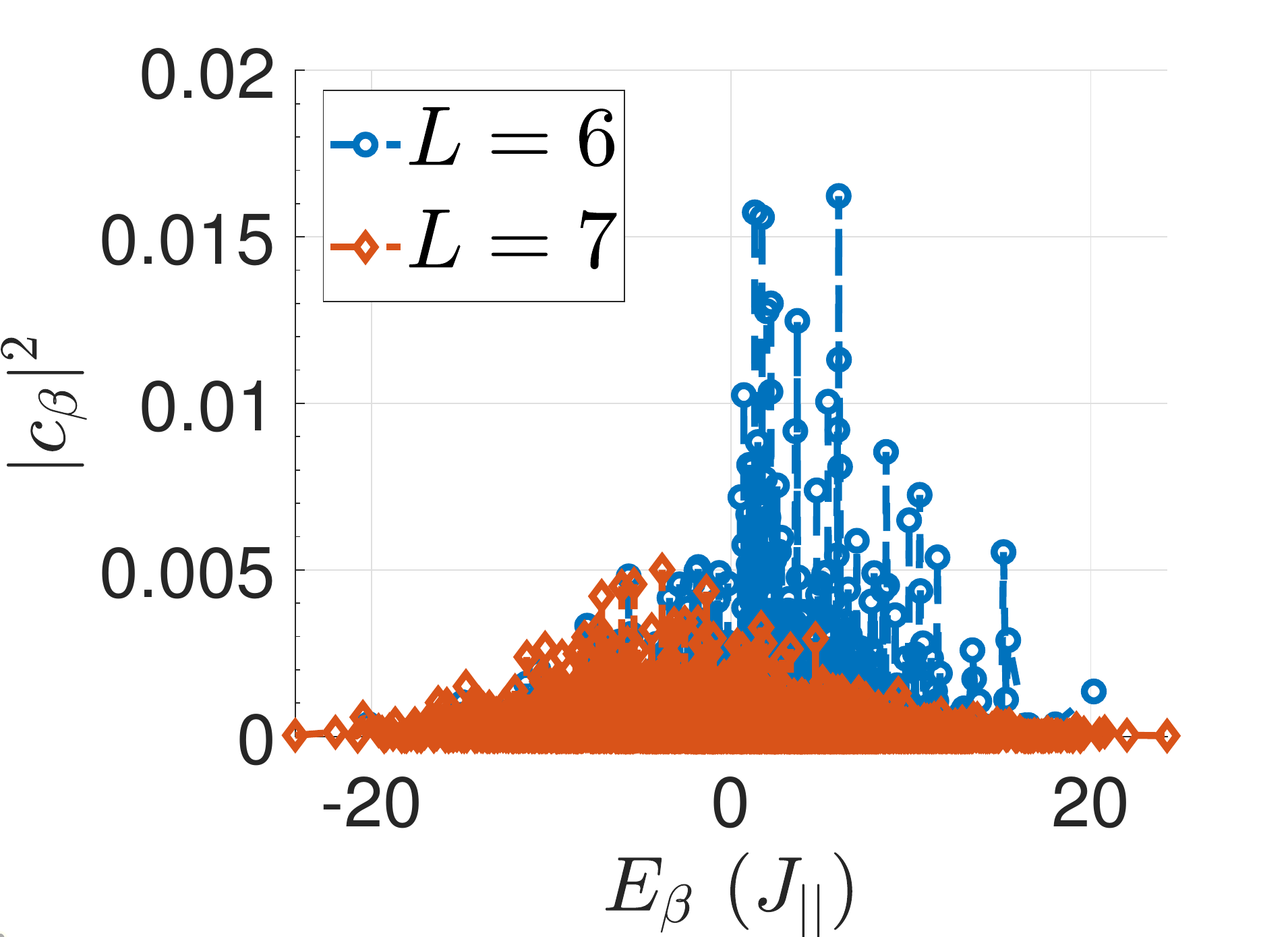}}\hfill \subfloat[]{\label{supFig2}\includegraphics[width=0.24\textwidth]{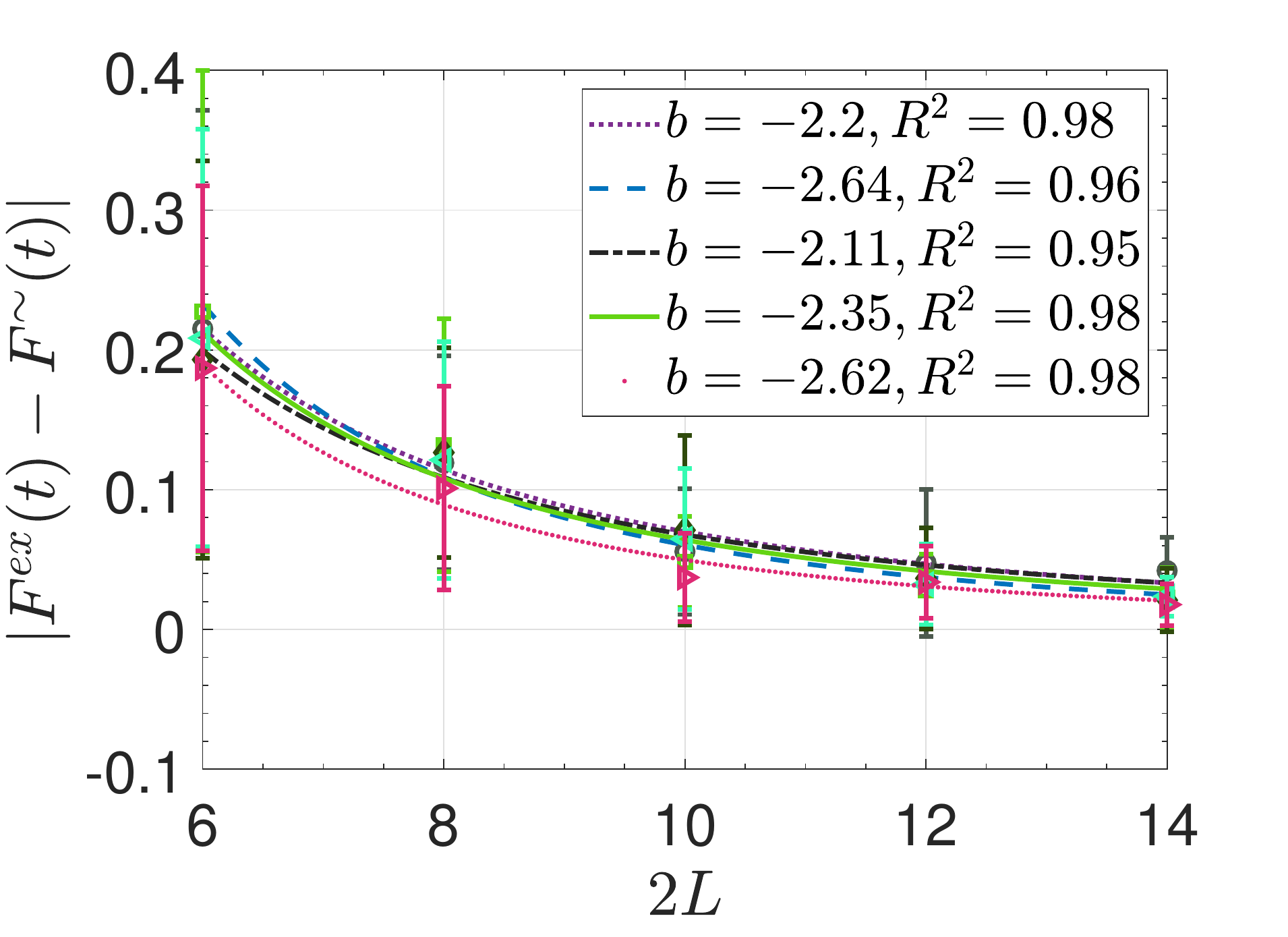}}\hfill 
\caption{(a) The EON (eigenstate occupation number) distributions $|c_{\beta}|^2$ with respect to eigenenergies $E_{\beta}$ for $L=6$ (blue) and $L=7$ (orange) sizes when only one Fock state is randomly set. (b) The scaling of the mean of the error $|F^{\text{ex}}(t)-\frac{1}{M}\sum_j F_j(t)|$ with the system size when we use only one randomly-sampled Fock state. Different curves are different random realizations with the legend showing the exponent of the corresponding power-law decay. The error bars stand for $1\sigma$ standard deviation around the mean of the error signal.}
\end{figure}
We give the plot that shows EON (eigenstate occupation numbers) distribution, $|c_{\beta}|^2$, for $L=6$ and $L=7$ in Fig.~\ref{supFig1} for a randomly-set initial Fock state. These distributions should be contrasted with a uniform distribution of an infinite-temperature initial state. Even though they are not uniform, they are still broad distributions which helps the approximation error to be bounded. As a result, we state that as long as the initial state has a broad distribution in the eigenbasis, the exact shape of the distribution is not significant. Hence such an initial state could be used to sufficiently approximate an infinite-temperature OTOC. 

Fig.~\ref{supFig2} shows the error $\epsilon_1$ scales as a power-law in the system size when only one Fock state is randomly-set. This figure focuses on five realizations that were given in the main text in logarithmic scale. Here we plot the data in linear scale to also demonstrate the error bars. The error bars stand for 1$\sigma$ deviation around the mean of the error signal in time. Note that the error bars increasingly become smaller as the system size increases, meaning that our initial state approximation does not only work better on average but also throughout the simulation time. 

Finally we provide the exact fitting expressions for the exponential and power-law scalings of the mean error in the sampling ratio $M/N$. The exponential scaling parameters are, $a=0.1218$, $R^2=0.9134$ ($N=3$), $a=0.043$, $R^2=0.933$ ($N=4$), $a=0.0132$, $R^2=0.884$ ($N=5$) and $a=0.004$, $R^2=0.962$ ($N=6$) with very close exponents $b\sim-2.5$. The power-law scaling parameters are, $a=0.0112$, $R^2=0.984$ ($N=3$), $a=0.0037$, $R^2=0.991$ ($N=4$), $a=8\times10^{-4}$, $R^2=0.945$ ($N=5$) and $a=3.4\times10^{-4}$, $R^2=0.981$ ($N=6$) with very close exponents $b\sim-0.5$.

\bibliographystyle{apsrev4-1}
%

\end{document}